\documentclass[final]{siamonline0516}

\usepackage{lipsum}
\usepackage{amsfonts}
\usepackage{graphicx}
\usepackage{epstopdf}
\usepackage{algorithmic}
\usepackage{subfigure}
\usepackage{hyperref}
\usepackage[utf8]{inputenc}
\usepackage[english]{babel}
\ifpdf
  \DeclareGraphicsExtensions{.eps,.pdf,.png,.jpg}
\else
  \DeclareGraphicsExtensions{.eps}
\fi

\newcommand{\TheTitle}{Joint Multichannel Deconvolution and Blind Source Separation} 
\newcommand{\TheAuthors}{Ming Jiang, Jérôme Bobin, Jean-Luc Starck}

\headers{\TheTitle}{\TheAuthors}

\title{{\TheTitle}}

\author{
  Ming Jiang\thanks{Service d'Astrophysique, CEA Saclay, France
    (\email{ming.jiang@cea.fr}, \email{jerome.bobin@cea.fr}, \email{jean-luc.starck@cea.fr}).}
  \and
  Jérôme Bobin\footnotemark[1]
  \and
  Jean-Luc Starck\footnotemark[1]
}

\usepackage{amsopn}

\DeclareMathOperator{\Th}{Th}

\ifpdf
\hypersetup{
  pdftitle={\TheTitle},
  pdfauthor={\TheAuthors}
}
\fi

\begin{document}
\maketitle
\overfullrule=0pt

\begin{abstract}
Blind Source Separation (BSS) is a challenging matrix factorization problem that plays a central role in multichannel imaging science. In a large number of applications, such as astrophysics, current unmixing methods are limited since real-world mixtures are generally affected by extra instrumental effects like blurring. Therefore, BSS has to be solved jointly with a deconvolution problem, which requires tackling a new inverse problem:  deconvolution BSS (DBSS). In this article, we introduce an innovative DBSS approach, called DecGMCA, based on sparse signal modeling and an efficient alternative projected least square algorithm. Numerical results demonstrate that the DecGMCA algorithm performs very well on simulations. It further highlights the importance of jointly solving BSS and deconvolution instead of considering these two problems independently. Furthermore, the performance of the proposed DecGMCA algorithm is demonstrated on simulated radio-interferometric data.
\end{abstract}

\section{Introduction}
\label{sec:intro}
In many imaging applications, such as astrophysics, the advent of multi-wavelength instruments mandates the development of advanced signal/image processing tools to extract relevant information. It is especially difficult to identify and extract the characteristic spectra of sources of interest when they are blindly mixed in the observations.

For this reason, multichannel data has generated interest in study of the Blind Source Separation (BSS) problem. In the BSS model, suppose we have $\text{N}_\text{c}$ channels, each channel $\lbrace\mathbf{x}_i\rbrace_{1 \leq i \leq \text{N}_\text{c}}$ delivers an observation which is the linear combination of $\text{N}_\text{s}$ sources and contaminated by the noise $\lbrace\mathbf{n}_i\rbrace_{1 \leq i \leq \text{N}_\text{c}}$. More precisely,
\begin{equation}
\forall i \in \lbrace 1,2,\cdots,\text{N}_\text{c}\rbrace, \quad \mathbf{x}_i=\sum_{j=1}^{\text{N}_\text{s}}A_{ij}\mathbf{s}_j + \mathbf{n}_i,
\label{eq:bssmodel}
\end{equation}
where the matrix $\mathbf{A}$ is the mixing matrix. BSS problems appear in a variety of applications, such as in astronomy~\cite{bss_astro}, neuroscience~\cite{bss_eeg,bss_neuro}, or medical imaging~\cite{bss_fMRI}. Various BSS methods have been proposed in the literature. Depending on the way the sources are separated, most BSS methods can be divided into two main families. The first family is based on a statistical approach, called Independent Component Analysis (ICA), such as FastICA~\cite{Hyvarinen1999} and its derivatives~\cite{Koldovsky2005}. These approaches assume that the sources are statistically independent and build upon statistical independence as a discriminative measure of diversity between the sources. However, they are generally not robust at low Signal to Noise Ratio (SNR). In the last decade, a second family of approaches, based on morphological diversity and sparsity, has blossomed. The pioneering work of Zibulevsky \textit{et al.}~\cite{Zibulevsky2001} introduced a sparsity prior to distinguish sources. The sparsity constraint can be imposed directly on the image pixel values or on its coefficients in a given sparse representation in a different space, the GMCA method proposed by Bobin \textit{et al.}~\cite{Bobin2007} and its derivative AMCA~\cite{Bobin2015} enforce the sparsity constraint in the sparse domain and employ an adaptive threshold strategy to extract sparse sources. It has been shown that these methods are more tolerant to non-independent sources and more robust to noisy data than ICA based approaches.\\

BSS alone is a complex inverse problem since it is a non-convex problem matrix factorization problem. This problem becomes even more challenging when the data are not fully sampled or blurred due to the Point Spread Function (PSF). For instance, in radio-astronomy instruments such as the future Square Kilometre Array (SKA~\cite{ska09}) provide an incomplete sampling in Fourier space, leading to an ill-conditioned system. Furthermore, the PSF is not identical for the different channels of the instrument, which degrades the observed image channels differently. Therefore, for such instruments, apart from the source separation problem, we also need to restore images from masking or blurring effects. Mathematically, the masking or the blurring can be modeled as a linear operator. Hence, we have to jointly solve both a deconvolution and a BSS problem, yielding a \textbf{Deconvolution Blind Source Separation (DBSS)} problem. 

Deconvolution BSS problems have not been extensively studied in the literature. To the best of our knowledge, solving BSS problems from incomplete measurements has only been investigated in the framework of Compressed Sensing (CS) in work by Kleinsteuber~\cite{Kleinsteuber2012}. The CS-BSS problem can be considered as a specific case of DBSS with the linear operation specialized in masking. However, the proposed approach only applies to Compressed Sensing measurements, which is a very specific case of the DBSS problem we investigate in this paper. In CS framework as well, the Compressive Source Separation (CSS) method proposed in~\cite{Golbabaee2013} processes the source separation of hyperspectral data, but under the assumption that the mixing matrix is known. In the framework of ICA, the DBSS algorithm can be recast as a special case of BSS from convolutive mixture models \cite{Douglas2005,Kokkinakis2006,Tonazzini2010}. However, the methods that have been introduced to unmix convolutive mixtures provide an estimate of the mixing matrix but not a joint estimation with the sources. These methods are limited to well-conditioned convolution operators, which excludes the ill-posed convolution operators we consider in this paper.

\subsection*{Contribution}
In this paper, we introduce a novel sparsity-based BSS algorithm that jointly addresses blind source separation and deconvolution. The proposed algorithm, coined DecGMCA, allows tackling blind separation problems that involve linear, potentially ill-conditioned or ill-posed convolution operators. This includes incomplete measurements or blurring effect.

This paper is organized as follows: we will present our DBSS model in \cref{sec:model}. In \cref{sec:method} we will introduce our deconvolution BSS method, called DecGMCA. Numerical experiments follow in \cref{sec:experiments1} and we will demonstrate the performance of our method. Then in \cref{sec:experiments2} we will apply our method to realistic interferometric data to give an illustration of the technique.

\section*{Notation}
Before moving to the problem setup, we will introduce the notation which will be used hereafter. Matrices and vectors are denoted in boldface uppercase letters and lowercase letters such as $\mathbf{X}$ and $\mathbf{x}$ respectively. The individual entry of a matrix or a vector is denoted in normal font such as $X_{i,j}$ or $x_{i}$. $||\mathbf{x}||$ denotes the length or $\ell_2$ norm of a vector $\mathbf{x}$. $\mathbf{X}^t$ denotes nonconjugate transpose of $\mathbf{X}$, while $\mathbf{X}^*$ denotes the conjugate transpose of $\mathbf{X}$. The Fourier transform of $\mathbf{X}$ is defined as $\mathbf{\hat{X}}$. For any matrix $\mathbf{X}$, $||\mathbf{X}||_p$ denotes the $p$-norm of the matrix. Specifically, $||\mathbf{X}||_F$ is called the Frobenius norm of the matrix. The identity matrix of size $n$ is denoted by $\mathbf{I}_n$. In addition, $\odot$ denotes the Hadamard piecewise product and $*$ denotes convolution. Finally, as for the proximal operation, $\mathbf{Y}=\Th_{\lambda}(\mathbf{X})$ returns the thresholding of the input $\mathbf{X}$ with threshold $\lambda$ (soft or hard thresholding is not specified, it will be clarified if needed).

More precisely, in terms of our DBSS problem, given $\text{N}_\text{s}$ sources of length $\text{N}_\text{p}$, sources are written as a stack of row vectors (two dimensional source will be aligned into row vector), denoted as $\mathbf{S}=(S_{i,j})_{\substack{1\leq i\leq \text{N}_\text{s}\\1\leq j\leq \text{N}_\text{p}}}=[\mathbf{s}_1^t,\mathbf{s}_2^t,\cdots,\mathbf{s}_{\text{N}_\text{p}}^t]^t$, therefore $\lbrace\mathbf{s}_i\rbrace_{1 \leq i \leq \text{N}_\text{p}}$ denotes the $i$-th source. In order to simplify the presentation of the model hereafter, the source matrix will also be written as concatenated column vectors such as $\mathbf{S}=[\mathbf{s}^1,\mathbf{s}^2,\cdots,\mathbf{s}^{\text{N}_\text{p}}]$, where $\lbrace\mathbf{s}^j\rbrace_{1 \leq j \leq \text{N}_\text{p}}$ is a column vector of all sources at position $j$. Assuming we have $\text{N}_\text{c}$ channels, the mixing matrix is written as a stack of row vectors such as $\mathbf{A}=(A_{i,j})_{\substack{1\leq i\leq \text{N}_\text{c}\\1\leq j\leq \text{N}_\text{s}}}=[\mathbf{a}_1^t,\mathbf{a}_2^t,\cdots,\mathbf{a}_{\text{N}_\text{c}}^t]^t$, where $\lbrace \mathbf{a}_i \rbrace_{1\leq i \leq \text{N}_\text{c}}$ is a row vector of the contribution of all sources at channel index $i$. The kernel $\mathbf{H}=(H_{i,j})_{\substack{1\leq i\leq \text{N}_\text{c}\\1\leq j\leq \text{N}_\text{p}}}$, which takes into account the masking or the blurring effect due to the PSF, is written as $\mathbf{H}=[\mathbf{h}_1^t,\mathbf{h}_2^t,\cdots,\mathbf{h}_{\text{N}_\text{c}}^t]^t$, where $\lbrace \mathbf{h}_i \rbrace_{1\leq i \leq \text{N}_\text{c}}$ is a row vector of the PSF at channel index $i$. Finally, the observation is denoted as $\mathbf{Y}=(Y_{\nu,k})_{\substack{1\leq \nu \leq \text{N}_\text{c}\\1\leq k \leq \text{N}_\text{p}}}=[\mathbf{y}_1^t,\mathbf{y}_2^t,\cdots,\mathbf{y}_{\text{N}_\text{c}}^t]^t$, where $\lbrace \mathbf{y}_i \rbrace_{1\leq i \leq \text{N}_\text{c}}$ is a row vector giving the observation at channel index $i$.

\section{The DBSS Problem}
\label{sec:model}
As shown in \Cref{eq:bssmodel}, we have $\text{N}_\text{c}$ channels available for the observation and each observation channel is assumed to be a mixture of $\text{N}_\text{s}$ sources (each source is of length $\text{N}_\text{p}$). The columns of mixing matrix $\mathbf{A}$ define the contribution of the sources in the mixture and are regarded as spectral signatures of the corresponding sources. We assume herein that the number of channels is greater than or equal to the number of sources: $\text{N}_\text{c} \geq \text{N}_\text{s}$ and $\mathbf{A}$ is a full-rank matrix. Besides the mixing stage, the observations are degraded by a linear operator $\mathbf{H}$:
\begin{enumerate}
\item[-] On the one hand, the data may be sub-sampled and this issue is related to the compressed sensing data. $\mathbf{H}$ can be therefore interpreted as a sub-sampling matrix or a mask.
\item[-] On the other hand, the data may be blurred by a PSF and $\mathbf{H}$ is a convolution operator.
\end{enumerate}
Moreover, the observed data are contaminated with the additive noise $\mathbf{N}$. Hence, the proposed imaging model can be summarized as follows:
\begin{equation}
\forall \nu\in\lbrace 1,2,\cdots,\text{N}_\text{c} \rbrace, \; \mathbf{y}_\nu=\mathbf{h}_\nu*\mathbf{x}_\nu=\mathbf{h}_\nu*\left(\sum_{j=1}^{\text{N}_\text{s}} A_{\nu,j}\mathbf{s}_j\right)+\mathbf{n}_\nu=\mathbf{h}_\nu*\left(\mathbf{a}_{\nu}\mathbf{S}\right)+\mathbf{n}_\nu.
\label{eq:imag}
\end{equation}

If we apply a Fourier transform on both sides of the above equation, our model can be more conveniently described in the Fourier domain. We denote $\hat{\mathbf{S}}_{i,\cdot}$ the Fourier transform of the $i$-th source (for a two dimensional source, a two dimensional Fourier transform is applied and the Fourier coefficients are aligned as a row vector). For simplicity, $\hat{\mathbf{S}}$ is defined as a stack of row vectors of the Fourier coefficients of all sources. Using the same convention, the Fourier transform of the observation, the kernel and the noise can be defined respectively as $\hat{\mathbf{Y}}$, $\hat{\mathbf{H}}$ and $\hat{\mathbf{N}}$. The matrix $\hat{\mathbf{S}}$ will be written as concatenated column vectors such as $\hat{\mathbf{S}}=[\hat{\mathbf{s}}^1,\hat{\mathbf{s}}^2,\cdots,\hat{\mathbf{s}}^{\text{N}_\text{p}}]$, or a stack of row vectors $[\hat{\mathbf{s}}_1^t,\hat{\mathbf{s}}_2^t,\cdots,\hat{\mathbf{s}}_{\text{N}_\text{s}}^t]^t$, while $\hat{\mathbf{Y}}=[\hat{\mathbf{y}}_1^t,\hat{\mathbf{y}}_2^t,\cdots,\hat{\mathbf{y}}_{\text{N}_\text{c}}^t]^t$, $\hat{\mathbf{H}}=[\hat{\mathbf{h}}_1^t,\hat{\mathbf{h}}_2^t,\cdots,\hat{\mathbf{h}}_{\text{N}_\text{c}}^t]^t$ and $\hat{\mathbf{N}}=[\hat{\mathbf{n}}_1^t,\hat{\mathbf{n}}_2^t,\cdots,\hat{\mathbf{n}}_{\text{N}_\text{c}}^t]^t$ are written as stacks of row vectors. Thus, in the Fourier domain, our model can be recast as follows:
\begin{equation}
\forall \nu\in\lbrace 1,2,\cdots,\text{N}_\text{c} \rbrace, \; \hat{\mathbf{y}}_\nu=\hat{\mathbf{h}}_\nu\odot\hat{\mathbf{x}}_\nu=\hat{\mathbf{h}}_\nu\odot\left(\sum_{j=1}^{\text{N}_\text{s}} A_{\nu,j}\hat{\mathbf{s}}_j\right)+\hat{\mathbf{n}}_\nu=\hat{\mathbf{h}}_\nu\odot\left(\mathbf{a}_{\nu}\hat{\mathbf{S}}\right)+\hat{\mathbf{n}}_\nu,
\label{eq:imag_FT}
\end{equation}
where $\odot$ denotes the Hadamard product. More precisely, at frequency $k$ of channel $\nu$, the entity of $\hat{\mathbf{Y}}$ satisfies:
\begin{equation}
\hat{Y}_{\nu,k} = \hat{H}_{\nu,k}\mathbf{a}_{\nu}\hat{\mathbf{s}}^k + \hat{N}_{\nu,k}.
\label{eq:model_scale}
\end{equation}

This forward model applies to a large number of applications. For instance, in radioastronomy or in medicine, instruments such as a radio interferometer or a Magnetic Resonance Imaging (MRI) scanner actually measure Fourier components. The observations are sub-sampled or blurred during the data acquisition and sources of interest are mixed blindly. Therefore, blind source separation from degraded data has generated interest in both domains.

Sparsity has been shown to highly improve the separation of sources~\cite{Bobin2007}. We want to utilize this concept to facilitate the source separation. To solve the DBSS problem, we assume the $\text{N}_\text{s}$ sources forming the source matrix $\mathbf{S}$ are sparse in the dictionary $\mathbf{\Phi}$. Namely,
\begin{equation}
\forall i\in \lbrace 1,2,\cdots, \text{N}_\text{s}\rbrace;\quad \mathbf{s}_i=\boldsymbol\alpha_{i}\mathbf{\Phi},
\end{equation}
where $\mathbf{\Phi}$ is also called the $\textit{synthesis}$ operator, which reconstructs $\mathbf{s}_i$ by assembling coefficients $\boldsymbol{\alpha}_i$. Conversely, the $\textit{analysis}$ operator $\mathbf{\Phi}^t$ decomposes the source by a series of coefficients $\boldsymbol\alpha_i$ attached to atoms: $\boldsymbol\alpha_i=\mathbf{s}_i\mathbf{\Phi}^t$. In addition, the dictionary $\mathbf{\Phi}$ is supposed to be \textit{(bi-)orthogonal} in the above equation and the algorithm hereafter. If a dictionary is \textit{(bi-)orthogonal}, it satisfies $\mathbf{\Phi}\mathbf{\Phi}^t=\mathbf{\Phi}^t\mathbf{\Phi}=\mathbf{I}$. However, an $\textit{overcomplete}$ or a $\textit{redundant}$ dictionary allows for more degrees of freedom which helps signal/image restoration. Such a dictionary satisfies the exact reconstruction property $\mathbf{\Phi}^t\mathbf{\Phi}=\mathbf{I}$, but $\mathbf{\Phi}\mathbf{\Phi}^t$ is not guaranteed to be an identity matrix. Fortunately, for most redundant dictionaries, such as curvelets~\cite{starck2002}, $\mathbf{\Phi}\mathbf{\Phi}^t$ is diagonally dominant and such dictionaries can be considered as a good approximation of a tight frame. Although the demonstration and the algorithm hereafter are based on the (bi-)orthogonal dictionary, they are good approximations when tight frame dictionaries are used.

Therefore, under the sparsity constraint, our problem can be written in Lagrangian form as follows:
\begin{equation}
\min_{ \mathbf{S}, \mathbf{A}} \frac{1}{2}\sum\limits_\nu^{\text{N}_\text{c}}\sum\limits_k^{\text{N}_\text{p}}||\hat{Y}_{\nu,k} - \hat{H}_{\nu,k}\mathbf{a}_{\nu}\hat{\mathbf{s}}^k||_2^2 + \sum\limits_i^{\text{N}_\text{s}} \lambda_{i} ||\mathbf{s}_i\mathbf{\Phi}^t||_p,
\label{eq:opt}
\end{equation}
where the $\ell_p$ norm, which can be replaced by the $\ell_0$ norm or the $\ell_1$ norm, enforces the sparsity constraint in the dictionary $\mathbf{\Phi}$, while the quadratic term guarantees the data fidelity. Our goal is to recover the sources $\mathbf{S}$ and the mixing matrix $\mathbf{A}$ by jointly solving a deconvolution and a BSS problem. However, such problems are challenging, as BSS integrates deconvolution for multichannel data. First of all, the DBSS problem involves non-convex minimization, hence only a critical point can be expected. Then, the convolution kernel $\hat{\mathbf{H}}$ can be ill-conditioned or even rank deficient. As a consequence, the deconvolution can be unstable if not well regularized.

\section{DecGMCA: a sparse DBSS method}
\label{sec:method}
The GMCA framework proposed by Bobin \textit{et al.}~\cite{Bobin2007} is an efficient BSS method taking advantage of morphological diversity and sparsity in a transformed space. Compared to ICA-based methods, it has also been demonstrated to be more robust to noisy data. However, GMCA does not take deconvolution into account, which is limited in practical applications. Therefore, a more rigorous BSS method should be conceived for the DBSS problem.

In this section, we will firstly present several ingredients of our method before moving onto the whole algorithm. Then, we will discuss the initialization of the algorithm and the choice of parameters. We will discuss the convergence at the end of this section.

\subsection{Two-stage estimate}
\label{subsec:est}
As the original problem~\cref{eq:opt} is non-convex due to indeterminacy of the product $\mathbf{A}\hat{\mathbf{S}}$, reaching the global optimum can never be guaranteed. In the spirit of BCR~\cite{Tseng2001}, the product $\mathbf{A}\hat{\mathbf{S}}$ can be split into two variables $\mathbf{A}$ and $\hat{\mathbf{S}}$, which allows the original problem to be split into two alternating solvable convex sub-problems: estimate of $\mathbf{S}$ knowing $\mathbf{A}$
\begin{equation}
\min_{ \mathbf{S}} \frac{1}{2}\sum\limits_{\nu}^{\text{N}_\text{c}}\sum\limits_k^{\text{N}_\text{p}}||\hat{Y}_{\nu,k} - \hat{H}_{\nu,k}\mathbf{a}_{\nu}\hat{\mathbf{s}}^k||_2^2 + \sum\limits_i^{\text{N}_\text{s}} \lambda_{i} ||\mathbf{s}_{i}\mathbf{\Phi}^t||_p,
\label{eq:decgmca_S}
\end{equation}
and estimate of $\mathbf{A}$ knowing $\mathbf{S}$
\begin{equation}
\min_{\mathbf{A}} \frac{1}{2}\sum\limits_{\nu}^{\text{N}_\text{c}}\sum\limits_k^{\text{N}_\text{p}}||\hat{Y}_{\nu,k} - \hat{H}_{\nu,k}\mathbf{a}_{\nu}\hat{\mathbf{s}}^k||_2^2.
\label{eq:decgmca_A}
\end{equation}

\subsubsection{Estimate of S}
\label{subsubsec:estS}
Problem \cref{eq:decgmca_S} is convex but does not generally admit an explicit solution. To compute its minimizer requires resorting to iterative algorithms such as proximal algorithms ({\it e.g.} FISTA~\cite{FISTA}, Condat-Vu splitting method (\cite{condat2013,vu2013}) to only name two). However, these methods are very computationally demanding. In most cases the least squares method is sufficient to have a computationnally cheap rough estimate of the sources. Therefore, in the spirit of the GMCA algorithm, we will employ a projected least-squares estimation strategy. Assuming $f\left(\mathbf{a}_\nu,\hat{\mathbf{s}}^k\right) = \frac{1}{2}\sum\limits_{\nu}^{\text{N}_\text{c}}\sum\limits_k^{\text{N}_\text{p}}||\hat{Y}_{\nu,k} - \hat{H}_{\nu,k}\mathbf{a}_{\nu}\hat{\mathbf{s}}^k||_2^2$. In order to estimate $\hat{\mathbf{S}}$ with respect to $\mathbf{A}$, we should let the deviation of $f\left(\mathbf{a}_\nu,\hat{\mathbf{s}}^k\right)$ of $\hat{\mathbf{s}}^k$ vanish:
$\frac{\partial f\left(\mathbf{a}_\nu,\hat{\mathbf{s}}^k\right)}{\partial \hat{\mathbf{s}}^k} = 0$.
\begin{subequations}
In other words,
\begin{align}
\frac{\partial f\left(\mathbf{a}_\nu,\hat{\mathbf{s}}^k\right)}{\partial \hat{\mathbf{s}}^k} &= \sum\limits_{\nu}^{\text{N}_\text{c}}\sum\limits_{k}^{\text{N}_\text{p}} \left(\hat{H}_{\nu,k}\mathbf{a}_\nu\right)^t \left(\hat{Y}_{\nu,k} - \hat{H}_{\nu,k} \mathbf{a}_\nu \hat{\mathbf{s}}^k\right)\\
&= \sum\limits_{\nu}^{\text{N}_\text{c}}\sum\limits_{k}^{\text{N}_\text{p}} \hat{H}_{\nu,k} \hat{Y}_{\nu,k} \mathbf{a}_\nu^t - \sum\limits_{\nu}^{\text{N}_\text{c}}\sum\limits_{k}^{\text{N}_\text{p}} \left(\hat{H}_{\nu,k}\mathbf{a}_\nu\right)^t \left(\hat{H}_{\nu,k}\mathbf{a}_\nu\right) \hat{\mathbf{s}}^k = 0. 
\end{align}
\end{subequations}

For each position $k$, noticing that $\hat{H}_{\nu,k}$ is a scalar, we have
\begin{subequations}
\begin{align}
\sum\limits_{\nu}^{\text{N}_\text{c}}\hat{H}_{\nu,k} \hat{Y}_{\nu,k} \mathbf{a}_\nu^t - \left(\sum\limits_{\nu}^{\text{N}_\text{c}} \left(\hat{H}_{\nu,k}\mathbf{a}_\nu\right)^t \hat{H}_{\nu,k}\mathbf{a}_\nu\right) \hat{\mathbf{s}}^k &= 0 \\
\Rightarrow \hat{\mathbf{s}}^k = \left (\sum\limits_{\nu}^{\text{N}_\text{c}} (\hat{H}_{\nu,k}\mathbf{a}_\nu)^t (\hat{H}_{\nu,k}\mathbf{a}_\nu)\right )^{-1}\sum\limits_{\nu}^{\text{N}_\text{c}}\hat{H}_{\nu,k} \hat{Y}_{\nu,k} \mathbf{a}_\nu^t. 
\label{eq:expS}
\end{align}
\end{subequations}

In this article, the convolution kernels $\hat{\mathbf{H}}$ can be ill-conditioned or rank deficient. In this setting, the least-square estimate is either not defined if the inverse of the kernel is unbounded or highly unstable with an amplified level of noise. Therefore, we propose resorting to a Tikhonov regularization of the least-square estimate in Fourier space to stabilize the multichannel deconvolution step: 
\begin{equation}
\hat{\mathbf{s}}^k = \left(\sum\limits_{\nu}^{\text{N}_\text{c}} \left(\hat{H}_{\nu,k}\mathbf{a}_\nu\right)^t \left(\hat{H}_{\nu,k}\mathbf{a}_\nu\right)+\epsilon^\prime\mathbf{I}_{\text{N}_\text{s}}\right)^{-1}\sum\limits_{\nu}^{\text{N}_\text{c}}\hat{H}_{\nu,k} \hat{Y}_{\nu,k} \mathbf{a}_\nu^t,
\label{eq:regS}
\end{equation}
where $\mathbf{I}_{\text{N}_\text{s}}$ is an identity matrix of size $\text{N}_\text{s}$ with $\text{N}_\text{s}$ sources. $\epsilon^\prime\mathbf{I}_{\text{N}_\text{s}}$ is a regularization term that controls the condition number of the system. Since the condition number is dependent on the Fourier frequency $k$ (as our working space is Fourier space, $k$ corresponds to the frequency in Fourier space), denoting $\mathbf{P}(k)=\sum\limits_{\nu}^{\text{N}_\text{c}} \left(\hat{H}_{\nu,k}\mathbf{a}_\nu\right)^t \left(\hat{H}_{\nu,k}\mathbf{a}_\nu\right)$, we choose $\epsilon^\prime$ to be proportional to the spectral norm of matrix $\mathbf{P}$ such that $\epsilon^\prime(k)=\epsilon||\mathbf{P}(k)||_2$ with $\epsilon$ the regularization parameter to be discussed in \cref{subsubsec:reg}.

Unfortunately, the noise is not cleanly removed and artifacts are present after the above procedure. The next step consists in enforcing the sparsity of the sources in the wavelets space, which yields the following estimate of the sources:
\begin{equation}
\forall i\in\lbrace 1,2,\cdots,\text{N}_\text{s}\rbrace;\quad \mathbf{s}_{i} = \left(\Th_{\lambda_{i}}\left(\mathbf{s}_{i}\mathbf{\Phi}^t\right)\right)\mathbf{\Phi},
\label{eq:DecGMCA_rec_S}
\end{equation}
where $\Th_{\lambda_i}(\cdot)$ denotes the thresholding operation that will be discussed in \cref{subsubsec:th}. Besides, as mentioned before, $\mathbf{\Phi}$ is a (bi-)orthogonal dictionary during the demonstration.

In summary, equipped with the wavelet shrinkage, the multichannel hybrid Fourier-wavelet regularized deconvolution presented above performs regularization in both the Fourier and wavelet spaces and it can be interpreted as a multichannel extension of the ForWaRD deconvolution method~\cite{Neelamani2004}.

Using such a projected regularized least-square source estimator is motivated by its lower computational cost. If this procedure provides a more robust separation process, it does not provide an optimal estimate of the sources. Consequently, in the last iteration, the problem is properly solved so as to provide a very clean estimate of $\mathbf{S}$. Solving \cref{eq:decgmca_S} is then carried out with a minimization method based on the Condat-Vu splitting method (\cite{condat2013,vu2013}).

\subsubsection{Estimate of A}
\label{subsubsec:estA}
Similarly, we derive the mixing matrix $\mathbf{A}$ from $\hat{\mathbf{S}}$ by vanishing the deviation of $f\left(\mathbf{a}_\nu,\hat{\mathbf{s}}^k\right)$ of $\mathbf{a}_\nu$:
$\frac{\partial f\left(\mathbf{a}_\nu,\hat{\mathbf{s}}^k\right)}{\partial \mathbf{a}_\nu} = 0$. Having noticed that $\hat{\mathbf{s}}^k$ is complex valued, we obtain:
\begin{subequations}
\begin{align}
\frac{\partial f\left(\mathbf{a}_\nu,\hat{\mathbf{s}}^k\right)}{\partial \mathbf{a}_\nu} &= \sum\limits_{\nu}^{\text{N}_\text{c}}\sum\limits_{k}^{\text{N}_\text{p}} \left(\hat{Y}_{\nu,k} - \hat{H}_{\nu,k} \mathbf{a}_\nu \hat{\mathbf{s}}^k\right)\left(\hat{H}_{\nu,k}\hat{\mathbf{s}}^k\right)^* \\
&= \sum\limits_{\nu}^{\text{N}_\text{c}}\sum\limits_{k}^{\text{N}_\text{p}} \hat{H}_{\nu,k}\hat{Y}_{\nu,k} \left(\hat{\mathbf{s}}^k\right)^* - \sum\limits_{\nu}^{\text{N}_\text{c}}\sum\limits_{k}^{\text{N}_\text{p}} \mathbf{a}_\nu \left(\hat{H}_{\nu,k} \hat{\mathbf{s}}^k\right) \left(\hat{H}_{\nu,k}\hat{\mathbf{s}}^k\right)^* = 0.
\end{align}
\end{subequations}

For each frequency channel $\nu$, noticing that $\hat{H}_{\nu,k}$ is a scalar, the final expression is given by
\begin{subequations}
\begin{align}
\sum\limits_{k}^{\text{N}_\text{p}}\hat{H}_{\nu,k} \hat{Y}_{\nu,k} \left(\hat{\mathbf{s}}^k\right)^* - \mathbf{a}_\nu \left(\sum\limits_{k}^{\text{N}_\text{p}} \left(\hat{H}_{\nu,k}\hat{\mathbf{s}}^k\right) \left(\hat{H}_{\nu,k}\hat{\mathbf{s}}^k\right)^*\right) &= 0 \\
\Rightarrow \mathbf{a}_\nu = \left(\sum\limits_{k}^{\text{N}_\text{p}}\hat{H}_{\nu,k} \hat{Y}_{\nu,k} \left(\hat{\mathbf{s}}^k\right)^*\right)\left(\sum\limits_{k}^{\text{N}_\text{p}} \left(\hat{H}_{\nu,k}\hat{\mathbf{s}}^k\right)\left(\hat{H}_{\nu,k}\hat{\mathbf{s}}^k\right)^*\right)^{-1}. 
\label{eq:expA}
\end{align}
\end{subequations}

Since $\text{N}_\text{p} \gg \text{N}_\text{c}$, the least-square term of update $\mathbf{A}$, which involves summations over all the $\text{N}_\text{p}$ samples at channel $\nu$, is not rank deficient and robust to be inverted. The estimate of $\mathbf{A}$ does not require an extra regularization parameter. As the $\ell_p$-norm constraint imposes a minimal norm of $\mathbf{S}$, the global optimization problem may diverge to an unexpected solution as $\mathbf{S}=\mathbf{0}$ and $\mathbf{A}=\mathbf{\infty}$. Therefore, it is necessary to renormalize the columns of $\mathbf{A}$ as unit vectors before updating next $\mathbf{S}$:
\begin{equation}
\forall j \in\lbrace 1,2,\cdots,\text{N}_\text{s}\rbrace;\quad \bar{\mathbf{a}}^j=\frac{\mathbf{a}^j}{||\mathbf{a}^j||}.
\label{eq:normA}
\end{equation}

\subsection{DecGMCA algorithm}
\label{subsec:DecGMCA}
Assembling the two-stage estimates, we summarize our Deconvolved-GMCA (DecGMCA) algorithm presented in \cref{alg:DecGMCA}: 

\begin{algorithm}[ht!]
\caption{Deconvolved-GMCA (DecGMCA)}
\label{alg:DecGMCA}
\begin{algorithmic}[1]
\STATE{\textbf{Input}: Observation $\hat{\mathbf{Y}}$, operator $\hat{\mathbf{H}}$, maximum iterations $\text{N}_\text{i}$, $\epsilon^{(0)}$}
\STATE{Initialize $\mathbf{A}^{(0)}$}
\FOR{$i=1,\ldots,\text{N}_\text{i}$}
 \STATE{$\ast\ast\ast\ast\ast\ast\ast\ast\ast\ast\ast\ast\ast\ast\ast\ast\ast\ast\ast$ Estimating $\mathbf{S}$ $\ast\ast\ast\ast\ast\ast\ast\ast\ast\ast\ast\ast\ast\ast\ast\ast\ast\ast\ast\ast\ast$}
	\FOR{$k=1,\ldots,\text{N}_\text{p}$}
		\STATE{$\bullet$ Compute the current $\hat{\mathbf{s}}^k$ with respect to the current estimate of $\mathbf{A}^{(i)}$:}
		\STATE{$(\hat{\mathbf{s}}^k)^{(i)} = \left(\sum\limits_{\nu}^{\text{N}_\text{c}} \left(\hat{H}_{\nu,k}\mathbf{a}_\nu^{(i)}\right)^t \left(\hat{H}_{\nu,k}\mathbf{a}_\nu^{(i)}\right)+\epsilon^\prime\mathbf{I}_N\right)^{-1}\sum\limits_{\nu}^{\text{N}_\text{c}}\hat{H}_{\nu,k} \hat{Y}_{\nu,k} (\mathbf{a}_\nu^{(i)})^t$}
	\ENDFOR
	\STATE{$\bullet$ Obtain sources in image space by inversing FFT:}
	\STATE{$\mathbf{S}^{(i)}=\text{Re}(\text{FT}^{-1}(\hat{\mathbf{S}}^{(i)}))$}
	\STATE{$\ast\ast\ast\ast\ast\ast\ast\ast\ast\ast\ast\ast\ast\ast\ast\ast\ast\ast\ast$ Sparse thresholding $\ast\ast\ast\ast\ast\ast\ast\ast\ast\ast\ast\ast\ast\ast\ast\ast\ast\ast\ast\ast\ast$}
	\FOR{$j=1,\ldots,\text{N}_\text{s}$}
		\STATE{$\bullet$ Apply sparsity prior in wavelet space and estimate the current coefficients by thresholding:}
		\STATE{$\boldsymbol\alpha_j=\Th_{\lambda_j^{(i)}}(\mathbf{s}_{j}^{(i)}\mathbf{\Phi}^t)$}
		\STATE{$\bullet$ Obtain the new estimate of $\mathbf{S}$ by reconstructing treated coefficients}
		\STATE{$\mathbf{s}_{j}^{(i)}=\boldsymbol\alpha_{j}\mathbf{\Phi}$}
	\ENDFOR
	\STATE{$\bullet$ Obtain sources in Fourier space by FFT:}
	\STATE{$\hat{\mathbf{S}}^{(i)}=\text{FT}(\mathbf{S}^{(i)})$}
	\STATE{$\ast\ast\ast\ast\ast\ast\ast\ast\ast\ast\ast\ast\ast\ast\ast\ast\ast\ast\ast$ Estimating $\mathbf{A}$ $\ast\ast\ast\ast\ast\ast\ast\ast\ast\ast\ast\ast\ast\ast\ast\ast\ast\ast\ast\ast\ast$}
	\FOR{$\nu=1,\ldots,\text{N}_\text{c}$}	
		\STATE{$\bullet$ Compute the current $\mathbf{a}_\nu$ with respect to the current estimate of $\mathbf{S}^{(i)}$:}
		\STATE{$\mathbf{a}_\nu^{(i)} = \left(\sum\limits_{k}^{\text{N}_\text{p}}\hat{H}_{\nu,k} \hat{Y}_{\nu,k} \left((\hat{\mathbf{s}}^k)^{(i)}\right)^*\right)\left(\sum\limits_{k}^{\text{N}_\text{c}} \left(\hat{H}_{\nu,k}(\hat{\mathbf{s}}^k)^{(i)}\right)\left(\hat{H}_{\nu,k}(\hat{\mathbf{s}}^k)^{(i)}\right)^*\right)^{-1}$}
	\ENDFOR
	\STATE{$\bullet$ Update the threshold $\lambda$ and $\epsilon$}
\ENDFOR
\STATE{$\ast\ast\ast\ast\ast\ast\ast\ast\ast\ast\ast\ast\ast\ast\ast\ast\ast\ast\ast\ast$ Ameliorating $\mathbf{S}$ $\ast\ast\ast\ast\ast\ast\ast\ast\ast\ast\ast\ast\ast\ast\ast\ast\ast\ast\ast\ast\ast\ast$}
\STATE\label{algo:line28}{Solve \Cref{eq:decgmca_S} with respect to $\mathbf{A}$ using proximal methods}
\RETURN $\mathbf{A},\mathbf{S}$
\end{algorithmic}
\end{algorithm}

\subsubsection{Initialization}
\label{subsubsec:init}
For the initialization of $\mathbf{A}$, we can simply take a random value as the first guess. Apart from random initialization, we can also utilize different strategies for the initialization following the specific form of the data:
\begin{enumerate}
\item[-] If the data are not sub-sampled, we can utilize SVD decomposition to help the initialization. Due to the size of $\hat{\mathbf{Y}}(\text{N}_\text{c}\ll\text{N}_\text{p})$, we perform a thin SVD decomposition such that $\hat{\mathbf{Y}}=\mathbf{U}_{\text{N}_\text{c}}\mathbf{\Sigma}_{\text{N}_\text{c}}\mathbf{V}^*$, the matrix $\mathbf{U}$ is thus of size $\text{N}_\text{c}\times\text{N}_\text{c}$, $\mathbf{\Sigma}_{\text{N}_\text{c}}$ is $\text{N}_\text{c}\times\text{N}_\text{c}$ diagonal matrix and $\mathbf{V}$ is of size $\text{N}_\text{c}\times\text{N}_\text{p}$. Then, the first guess of $\mathbf{A}$ is set to the first $\text{N}_\text{s}$ normalized columns of $\mathbf{U}$.
\item[-] If the data are sub-sampled, the discontinuity effect of the data affects the SVD initialization. In order to reduce such discontinuity, we perform a matrix completion scheme using the SVT algorithm~\cite{Cai2010} before the initialization:
\begin{equation}
\min ||\hat{\mathbf{X}}||_* \quad \text{s.t.}~||\hat{\mathbf{Y}}-\mathbf{H}\hat{\mathbf{X}}||_F^2 < err,
\label{eq:MC}
\end{equation}
where $||\cdot||_*$ denotes the nuclear norm. 
\end{enumerate}

\subsubsection{Regularization parameters}
\label{subsubsec:reg}
As shown in \Cref{eq:regS}, the Tikhonov term is of great importance to regularize the deconvolution procedure. Therefore, this section concerns the choice of the regularization parameter $\epsilon$. Recalling that $\epsilon^\prime=\epsilon||\mathbf{P}(k)||$, where $\mathbf{P}(k)=\sum\limits_{\nu}^{\text{N}_\text{c}} \left(\left(\hat{H}_{\nu,k}\mathbf{a}_\nu\right)^t \left(\hat{H}_{\nu,k}\mathbf{a}_\nu\right)\right)$, the Tikhonov term $\epsilon^\prime \mathbf{I}_{\text{N}_\text{s}}$ will control the condition number. Intuitively, a larger $\epsilon$ makes the system more regularized, but the detailed information will be smooth, yielding a loss of precision. In contrast, a smaller $\epsilon$ can conserve more details, but the system will not be sufficiently regularized and the deconvolution will be unstable.

%
%

Therefore, the parameter $\epsilon$ is fixed so as to provide a trade-off between precision and stability. During the first iterations, the source $\mathbf{S}$ is not well estimated and the estimation is vulnerable to the amplified noise. Thus, we apply a large $\epsilon$ to mainly regularize the ill-conditioned system and ensure that the solution will not get stuck in a local optimum. As the algorithm goes on, the sources $\mathbf{S}$ tend to converge towards a more stable solution. Then, we decrease $\epsilon$ to improve the estimate. However, $\epsilon$ can never be decreased to zero as zero regularization will make the estimate unstable again. In practice, $\epsilon$ decays linearly or exponentially from $10^{-1}$ to a very small non-zero value, for example $10^{-3}$. Besides, since the choice of final $\epsilon$ is dependent on the global condition number of the system and the tolerance of the precision, it should be adapted to the specific case in practice.

\subsubsection{Thresholding strategy}
\label{subsubsec:th}
We didn't specify the $\ell_p$ norm in the optimization problem~\cref{eq:opt}. Indeed, the $\ell_p$ norm can be either $\ell_0$ or $\ell_1$. The $\ell_0$ norm problem using hard-thresholding gives an exact sparse solution, while the $\ell_1$ norm problem using soft-thresholding, leading to a convex sub-problem, can be regarded as a relaxation of the $\ell_0$ norm. Nevertheless, hard-thresholding often converges to a better result in practice as it does not produce bias. Therefore, the sparsity constraint is written as an $\ell_0$ norm regularizer instead of an $\ell_1$ norm.

The sparsity parameters $\lbrace\lambda_i\rbrace_{1\leq i \leq \text{N}_\text{s}}$ can be implicitly interpreted as thresholds in \Cref{eq:DecGMCA_rec_S}. In addition, the choice of thresholds $\lbrace\lambda_i\rbrace_{1\leq i \leq \text{N}_\text{s}}$ is a vital point in the source separation process. The DecGMCA algorithm utilizes an adapted thresholding strategy. The initial thresholds are set to high values so that the most significant features of the sources can be extracted to facilitate source separation. In addition, the high thresholds prevent the algorithm from being trapped on local optima. When the most discriminant features of the sources are extracted following the high thresholds, the sources are separated with high probability. Then, to retrieve more detailed information about the sources, the thresholds decrease towards the final values. The final thresholds can be chosen as $\tau\sigma_i$ with $\sigma_i$ the standard deviation of noise of the $i$-th source. In practice, Median Absolute Deviation (MAD) is a robust empirical estimator for Gaussian noise. The value $\tau$ ranges between $2\sim 4$. In practice, there are many ways to chose the decreasing function of the threshold. We present our strategy of decreasing threshold called ``percentage decreasing threshold'', which is the most robust according to our tests. Assuming at iteration $i$, as for an ordered absolute wavelet coefficient set of the $j$-th source $|\boldsymbol\alpha_j|$, the current threshold is selected as the $p_j^{(i)}$-th element in $|\boldsymbol\alpha_j|$ such as $\lambda_j^{(i)}=|\boldsymbol\alpha_j|\left[p_j^{(i)}\right]$, where $p_j^{(i)}$ satisfies:
\begin{equation}
\forall j\in \lbrace 1,2,\cdots,\text{N}_\text{s} \rbrace;\quad p_j^{(i)} = \left(  \frac{(1-p_j^{(0)})(i-1)}{\text{N}_\text{i}-1}+p_j^{(0)}\right)\text{card}(|\boldsymbol\alpha_j|\geq \tau\sigma_j),
\label{eq:th}
\end{equation}
with $p_j^{(0)}$ the initial percentage (for example 5\%). Hence, $p_j^{(i)}$ increases linearly from $p_j^{(0)}$ to 100\%, or the thresholds decay until $\tau\sigma_j$.

\subsubsection{Convergence analysis}
\label{subsubsec:conv}
It is well-known that BSS problems are non-convex matrix factorization problems. Therefore, one can only expect to reach a critical point of the problem \cref{eq:opt}. Let us recall that the DecGMCA algorithm is built upon the Block Coordinate Relaxation minimization procedure where the two blocks are defined by the source matrix $\bf S$ and the mixing matrix $\bf A$. More generally, the DBSS problem can be described by the following generic formulation:
\begin{equation}
\min_{{\bf A},{\bf S}} f({\bf A}) + g({\bf S}) + h({\bf A}{\bf S})
\end{equation}
where $f$ stands for the $\ell_2$ ball constraint on the columns of the mixing matrix, $g$ for the $\ell_1$-norm penalization and $h$ for the data fidelity term. The convergence of BCR for this type of matrix factorization problem has been investigated by Paul Tseng in \cite{Tseng2001}. In this article, Tseng introduces conditions on the minimization problem that guarantees that the BCR alternate minimization procedure converges to a critical point of \cref{eq:opt}. We previously emphasized in \cite{Rapin_14_NMFwithSparse} that these convergence conditions apply to algorithms based on GMCA as long as the regularization parameters are kept fixed ({\it i.e.} the thresholds $\{\lambda_j\}$ and $\epsilon$). In the DecGMCA algorithms, these parameters evolve; this evolution is key to increase the robustness of the algorithm with respect to local stationary points. However, it has been noticed in \cite{Bobin2015} that these parameters tend to stabilize at the end of GMCA-like algorithms. In that case, the algorithm tends to be close to a regime where convergence is guaranteed by Tseng's paper. The same argument applies to the proposed DecGMCA algorithm.

\section{Numerical results on simulations}
\label{sec:experiments1}
In radioastronomy, the recent advent of giant ground-based radio interferometers brought improved angular, spectral and temporal resolutions. However, the interferometric data are sub-sampled and blurred in Fourier space, and the sources of interest are often mixed in multichannel interferometry imaging. Radio-interferometric data are the perfect candidate where a joint deconvolution and blind source separation problem needs to be solved. In the following, we will investigate the two following cases for the linear operator $\hat{\mathbf{H}}$:
\begin{enumerate}
\item[-] the data are sub-sampled because of a limited number of antennas of the interferometer. As for the compressed sensing data, the operator $\hat{\mathbf{H}}$, which is associated to the sub-sampling effect, can be regarded as a mask with value 1 for active data and 0 for inactive data;
\item[-] furthermore, the angular resolution of the interferometer is limited by its beamforming. In practice, the PSF of the interferometer is determined by its beam. Therefore, $\hat{\mathbf{H}}$, in more general case, can be considered as a PSF kernel, which can take any real value.
\end{enumerate}
Hence, depending on the form of the operator $\hat{\mathbf{H}}$, we will apply the DecGMCA algorithm to simulations corresponding to each case, namely a simulation on multichannel compressed sensing ({\it i.e. incomplete measurements}) and a simulation on multichannel deconvolution.

Firstly, we generate simulated but rather complex data so that we can easily launch Monte-Carlo tests with different parameter settings. The sources are generated as follows:
\begin{enumerate}
\item The mono-dimensional sources are K-sparse signals. The distribution of the active entries satisfy a Bernoulli process $\pi$ with parameter $\rho$ such that:
\begin{equation}
\mathbb{P}[\pi=1]=\rho,~\mathbb{P}[\pi=0]=1-\rho.
\end{equation}
\item Then the K-sparse sources are convolved with a Laplacian kernel with FWHM (Full Width at Half Maximum) equal to 20.
\end{enumerate}
Each source contains $\text{N}_\text{p}=4096$ samples and K is equal to 50. As mentioned in \cref{sec:model}, an overcomplete dictionary outperforms (bi-)orthogonal dictionary in terms of signal/image restoration. According to the form of the simulated sources or even the astrophysical sources in more general cases which are isotropic, the starlet is optimal to sparsely represent such sources~\cite{starck1994}.


Before moving to our numerical experiments, we firstly define the criteria that will be used to evaluate the performance of the algorithms:
\begin{enumerate}
\item[-] the criterion for the mixing matrix is defined as: $\Delta_\mathbf{A} = -\log_{10}\frac{||\mathbf{A}_{est}^\dagger \mathbf{A}_{ref}-\mathbf{I}_{\text{N}_\text{s}}||_1}{\text{N}_\text{s}^2}$, which is independent of the number of sources $\text{N}_\text{s}$. Intuitively, when the mixing matrix is perfectly estimated, $\Delta_\mathbf{A}=+\infty$. Thus, the larger $\Delta_\mathbf{A}$ is, the better the estimate of $\mathbf{A}$ will be.
\item[-] according to~\cite{Vincent2006}, the estimated sources can be evaluated by a global criterion Source to Distortion Ratio (SDR) in dB: $\text{SDR}=10\log_{10}\frac{||\mathbf{S}_{ref}||_2}{||\mathbf{S}_{est}-\mathbf{S}_{ref}||_2}$. The higher the SDR is, the better the estimate of $\mathbf{S}$ will be.
\end{enumerate}

The rest of this section is organized as follows: for each case of simulations, we firstly study the performance of DecGMCA in terms of the condition of $\hat{\mathbf{H}}$, the number of sources and the SNR, then we compare DecGMCA with other methods. If not specified, all criteria will be chosen as the median of 50 independent Mont-Carlo tests.

\begin{figure}[ht!]
	\begin{center}
	\begin{tabular}{c}
		\subfigure[Example of a real masked observation compared with the complete data. Remark: The raw data are in Fourier space, but they are transformed to pixel space for better visualization. The percentage of active data is 50\% and the SNR is 60 dB.]{\includegraphics[width=0.9\textwidth]{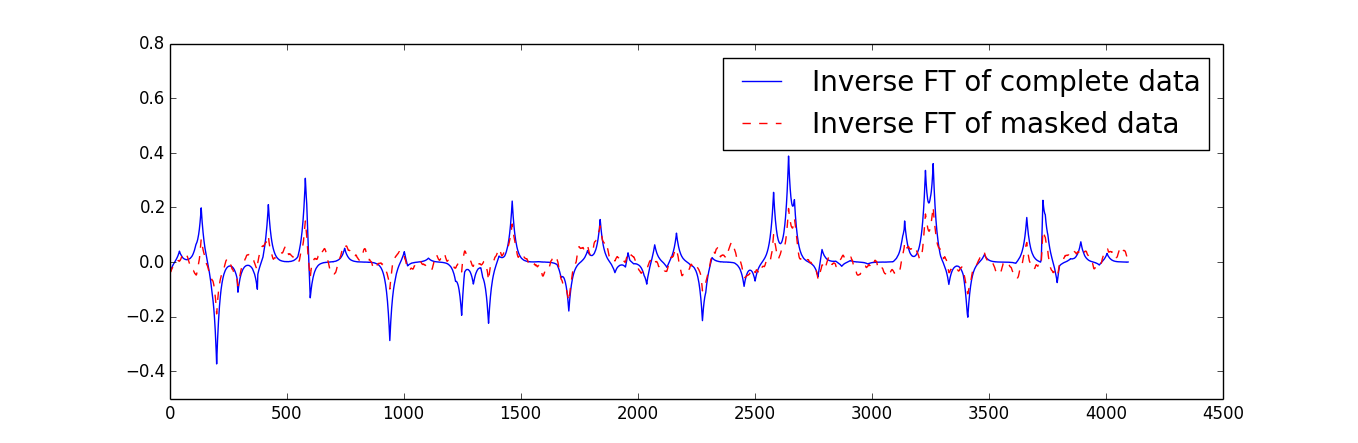}\label{fig:eg_sources:a}} \\
		\subfigure[Example of a recovered source from the above masked observations through DecGMCA superposed with the ground-truth.]{\includegraphics[width=0.9\textwidth]{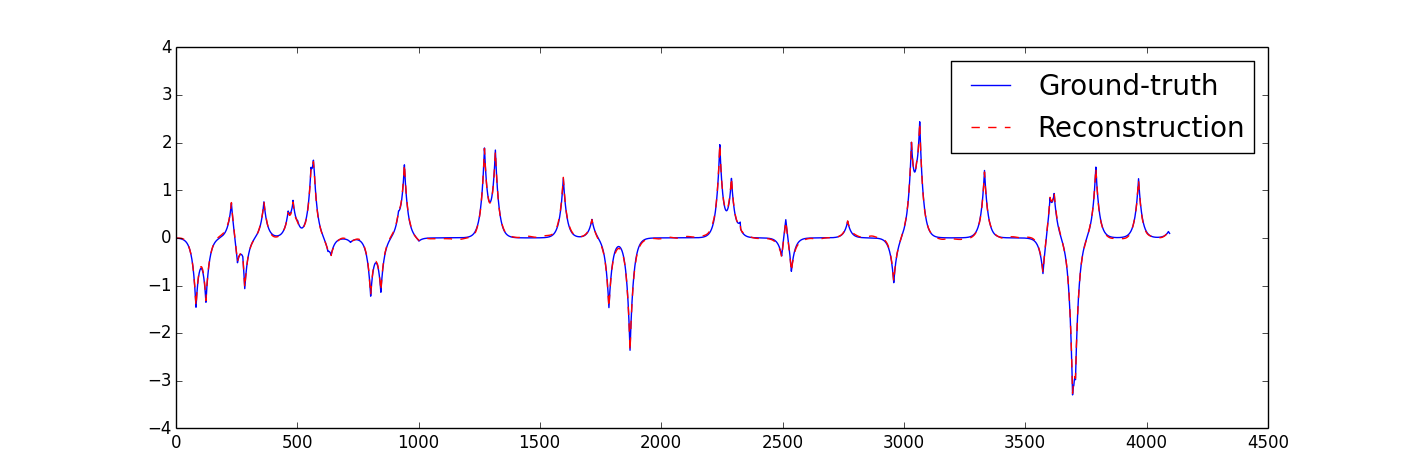}\label{fig:eg_sources:b}}	
	\end{tabular}
	\end{center}
\caption{Examples of the data and recovered sources superposed with ground-truth.}
\label{fig:eg_sources}
\end{figure}
\clearpage

\subsection{Multichannel compressed sensing and BSS simulation}
\label{subsec:csSimu}
First of all, to give a general idea of simulated data, \cref{fig:eg_sources:a} displays 1 mixture out of 20 of masked data (red dashed curve) compared with that of complete data (blue dashed curve). The percentage of active data is 50\% and the SNR is 60 dB. Since the data are masked in Fourier space, we perform an inverse Fourier transform to better visualize the data. To show the performance of our method DecGMCA, \cref{fig:eg_sources:b} displays 1 recovered source (red dashed curve) out of 2 superposed with the simulated ground-truth source (blue solid curve). We can observe that DecGMCA is able to recover all peaks and the estimated source is not biased. Besides, the noise is well removed in the estimated source. 

In this part we will discuss precisely the performance of DecGMCA on the multichannel compressed sensing data in terms of the sub-sampling effect, the number of sources and the SNR. The initial regularization parameter $\epsilon^{(0)}$ is set to be 1 and $\epsilon$ decreases to $10^{-3}$ for all the experiments in this section.

\subsubsection{On the study of DecGMCA}
\label{subsubsec:recCS}
\begin{enumerate}
\item[-] \textit{Sub-sampling effect:} To study the impact of the sub-sampling effect, the mask in Fourier space is varied from 10\% to 90\% where the value denotes the percentage of active data. The number of sources $\text{N}_\text{s}$ is fixed to 5 and the SNR is fixed to 60 dB. The number of observation channels $\text{N}_\text{c}$ is set to 5, 10 and 20. We applied our DecGMCA method and \cref{fig:resCS:a} shows an error bar plot of the criterion of $\mathbf{A}$ and SDR in terms of the mask. The blue, green and red curves represent results corresponding $\text{N}_\text{c}=$ 5, 10 and 20 respectively. In the figure, the first conclusion we can draw is that when the percentage of active data is below 20\%, DecGMCA performs badly in the sense of $\Delta_\mathbf{A}$ and SDR. This is due to the fact that when the mask is very ill-conditioned, we almost have no observation in the dataset so that we cannot correctly recover the mixing matrix and the sources. As the mask becomes better conditioned (the percentage of active data increases), we have more and more observations and we are expected to have better performance. It is interesting to notice that when the $\text{N}_\text{c}=5$, no matter which mask is used, the performance of DecGMCA is not good. The lack of performance is caused by the underdetermination of the system in the presence of mask when $\text{N}_\text{c}=\text{N}_\text{s}$. Besides, we can also observe that when $\text{N}_\text{c}=10$, DecGMCA performs well when the percentage of active data is above 50\%. It could be argued that though each of 10 channels is sub-sampled, statistically, the loss of observation can be fully compensated when the percentage of active data is 50\%, or in other words, we have on average 5 fully sampled channels. Considering $\text{N}_\text{c}=20$, we can see all criteria are stabilized when the percentage of active data is over 50\%. This is due to the fact that the DecGMCA reaches peak performance in this test scenario when the system is more and more well-conditioned.

\begin{figure}[h!]
	\begin{center}
	\begin{tabular}{c}
		\subfigure[Performance of DecGMCA in terms of the percentage of active data. The number of sources is 5 and the SNR is 60 dB. Abscissa: percentage of active data. Ordinate: criterion of mixing matrix for left figure and SDR for right figure.]{\includegraphics[width=0.8\textwidth]{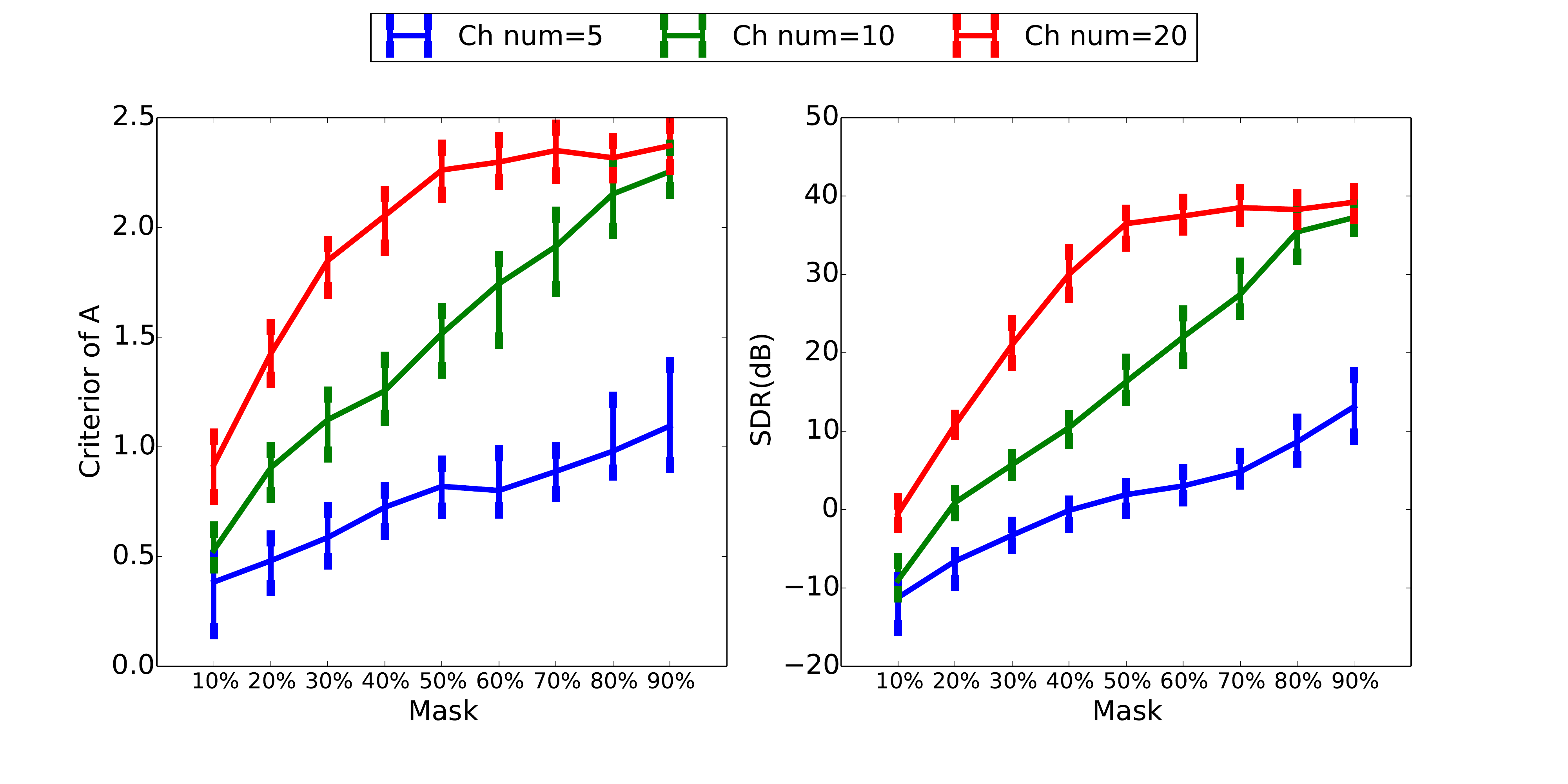}\label{fig:resCS:a}} \\ 
		\subfigure[Performance of DecGMCA in terms of the number of sources. The percentage of active data is 50\% and the SNR is 60 dB. Abscissa: number of sources. Ordinate: criterion of mixing matrix for left figure and SDR for right figure.]{\includegraphics[width=0.8\textwidth]{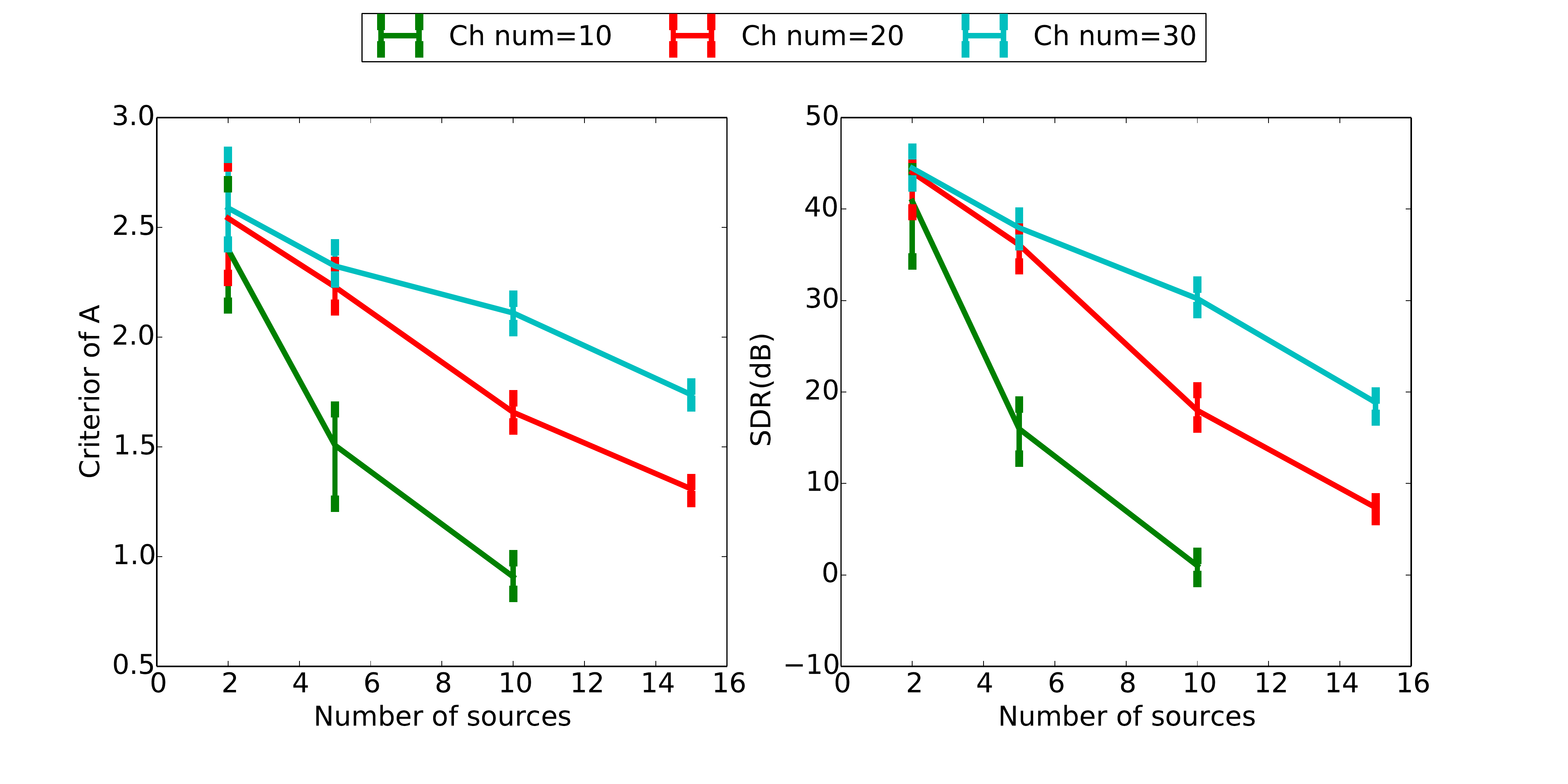}\label{fig:resCS:b}} \\
		\subfigure[Performance of DecGMCA in terms of SNR. The percentage of active data is 50\% and the number of sources is 5. Abscissa: SNR. Ordinate: criterion of mixing matrix for left figure and SDR for right figure.]{\includegraphics[width=0.8\textwidth]{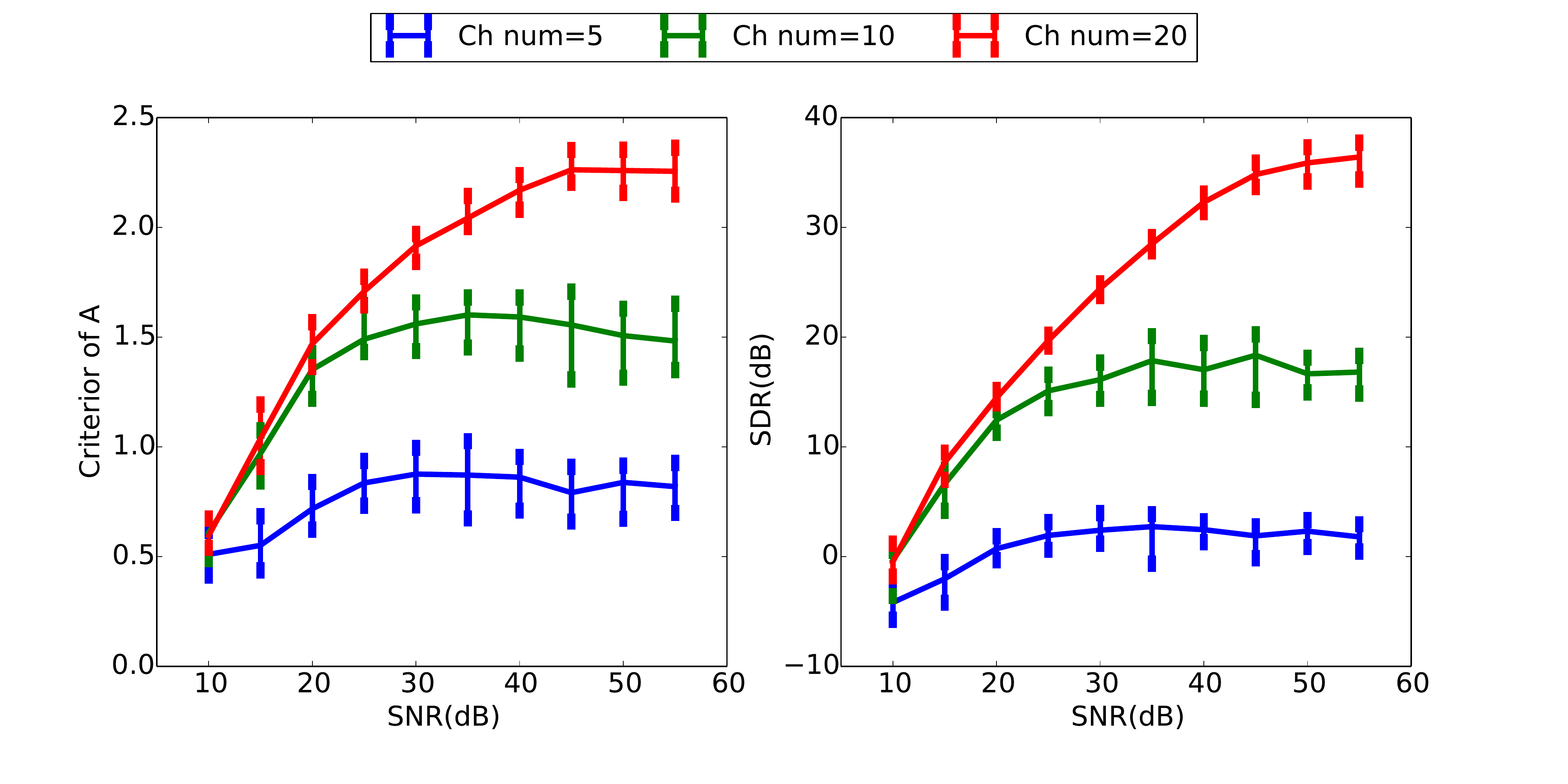}\label{fig:resCS:c}}
	\end{tabular}
	\end{center}
\caption{Multichannel compressed sensing simulation (1): study of DecGMCA. The parameters are the percentage of active data, the number of sources and the SNR from top to bottom. The curves are the medians of 50 realizations and the errorbars corresponds to 60\% of ordered data around the median. Blue curves, green curves, red curves and cyan curves represent the number of channels=5, 10, 20 and 30 respectively.}
\label{fig:resCS}
\end{figure}
\clearpage

\item[-] \textit{Number of sources:} The number of sources is also an important factor for the performance of DecGMCA. In this paragraph, $\text{N}_\text{s}$ is set to 2, 5, 10 and 15. The percentage of active data is fixed to 50\% and the SNR is fixed to 60 dB. $\text{N}_\text{c}$ is set to 10, 20 and 30. In \cref{fig:resCS:b},  when $\text{N}_\text{s}=2$, we can observe that the sub-sampling effect does not affect the performance. This is due to the fact that when the number of channels is sufficiently large compared to the number of sources, the loss of observation in one channel can be completely compensated by observations from other channels, which makes the matrix inversion easier and more accurate in the source estimate step (see \Cref{eq:regS}). Conversely, when $\text{N}_\text{s}=15$, noticing that the mask is 50\%, it is impossible to have good results when the number of channels is below 30. Thus, we can see in the figure that sources are not well recovered ($\text{SDR}<20$ dB) when $\text{N}_\text{c}$ is 20 or 30. Besides, given a fixed number of channels, the performance of DecGMCA decays as the number of sources increases, which is consistent with our expectation. Indeed, when the number of sources increases, the number of columns of the mixing matrix increases, which makes it more difficult to jointly recover the mixing matrix and reduce the masking effect.

\item[-] \textit{SNR:} The third parameter which affects the performance of DecGMCA is the SNR. In this paragraph, the SNR is varied from 10 dB to 55 dB. The percentage of active data is fixed to 50\% as well and $\text{N}_\text{s}$ is fixed to 5. $\text{N}_\text{c}$ is set to 5, 10 and 20. We can observe in \cref{fig:resCS:c} that as expected the performance gets worse as SNR decreases. Particularly, when $\text{N}_\text{c}=5$, irrespective of the SNR, DecGMCA performs poorly. This is owing to the fact when $\text{N}_\text{c}=5$ and the mask is 50\%, we are not able to successfully recover 5 sources (SDR is around 0 dB). Therefore, in this case, the number of channels instead of SNR is the bottleneck of the algorithm. When the SNR is below 25 dB for $\text{N}_\text{c}=10$ and 45 dB for $\text{N}_\text{c}=20$, we can see increasing the SNR significantly helps to improve the performance of DecGMCA. This can be argued that when the number of channels is no longer the restriction of the algorithm, the contamination of noise in the data becomes the dominant restriction of the performance of DecGMCA. When the SNR is high, it is easier to extract useful information to estimate sources. Thus, the sources are estimated with high precision. However, one should notice that the performance of DecGMCA cannot eternally grow along with the SNR. The reason for this is that we can already successfully extract information to estimate sources and an even higher SNR will not help us significantly. In this case, it is the number of channels that becomes the limiting factor for better estimating sources. Therefore, we can observe in the figure that the saturation points of the criteria of $\text{N}_\text{c}=20$ appears later than those of $\text{N}_\text{c}=10$.
\end{enumerate}

\begin{figure}[ht!]
	\begin{center}
	\begin{tabular}{c}
		\subfigure[Comparison between DecGMCA and MC+GMCA in terms of the percentage of active data. The number of sources is 5 and the SNR is 60 dB. Abscissa: percentage of active data. Ordinate: criterion of mixing matrix for left figure and SDR for right figure.]{\includegraphics[width=0.7\textwidth]{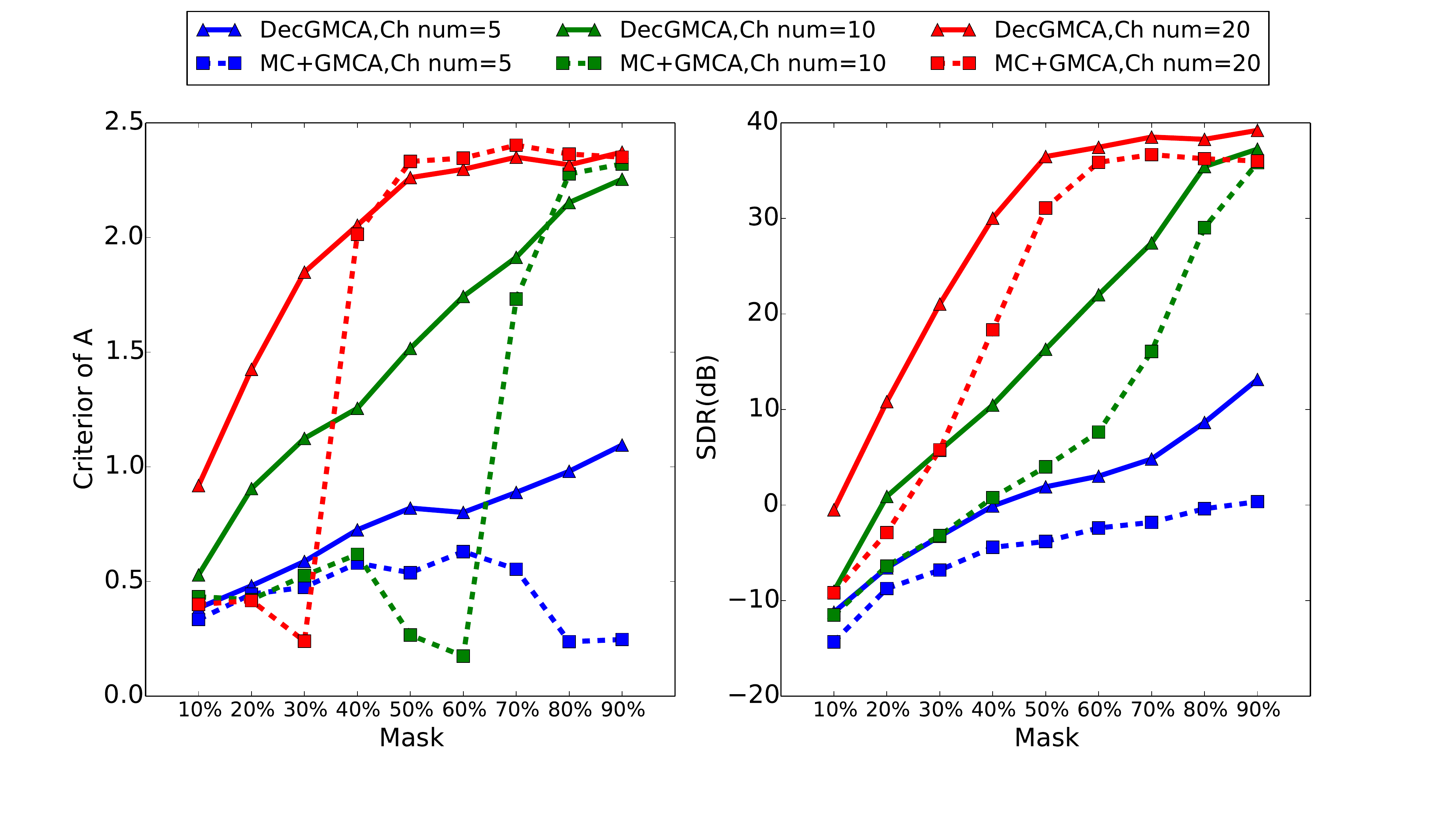}\label{fig:resCS_MC:a}} \\ 
		\subfigure[Comparison between DecGMCA and MC+GMCA in terms of the number of sources. The percentage of active data is 50\% and the SNR is 60 dB. Abscissa: number of sources. Ordinate: criterion of mixing matrix for left figure and SDR for right figure.]{\includegraphics[width=0.7\textwidth]{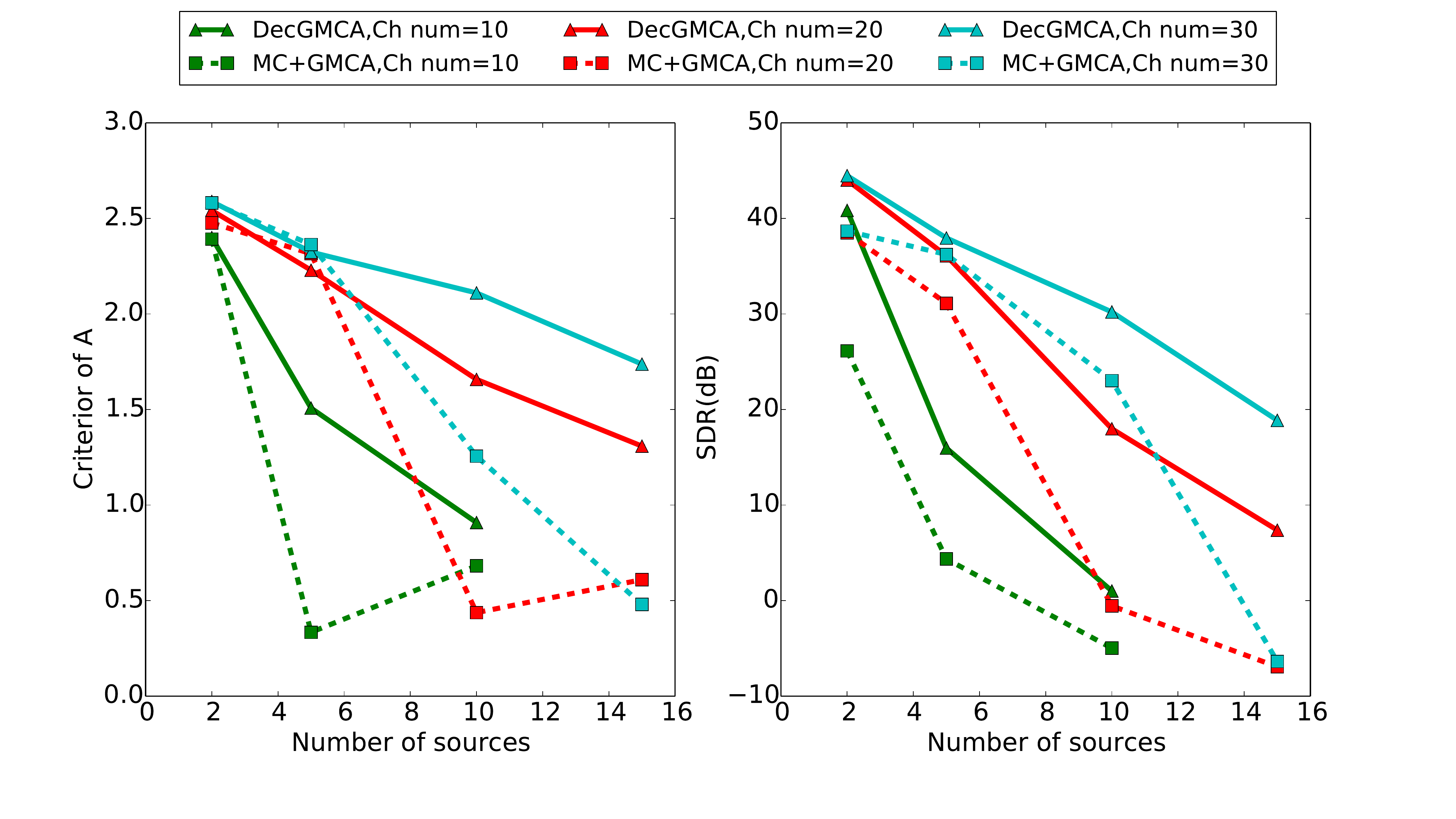}\label{fig:resCS_MC:b}} \\
		\subfigure[Comparison between DecGMCA and MC+GMCA in terms of SNR. The percentage of active data is 50\% and the number of sources is 5. Abscissa: SNR. Ordinate: criterion of mixing matrix for left figure and SDR for right figure.]{\includegraphics[width=0.7\textwidth]{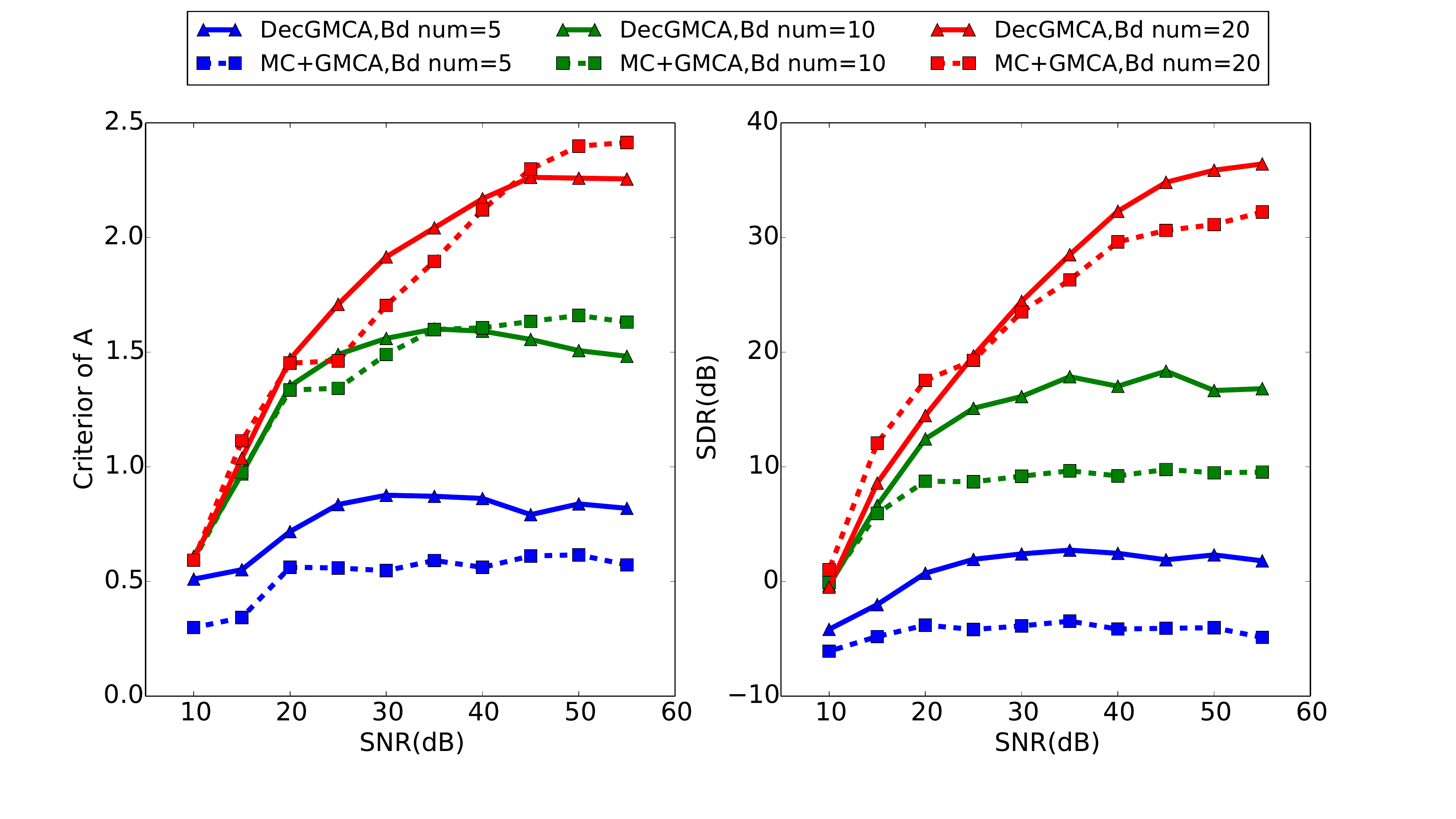}\label{fig:resCS_MC:c}}
	\end{tabular}
	\end{center}
\caption{Multichannel compressed sensing simulation (2): comparison between DecGMCA (joint deconvolution and BSS) and MC+GMCA (matrix completion followed by BSS). The parameters are the percentage of active data, the number of sources and the SNR from top to bottom. The curves are the medians of 50 realizations. Blue curves, green curves, red curves and cyan curves represent the number of channels=5, 10, 20 and 30 respectively.}
\label{fig:resCS_MC}
\end{figure} 
\clearpage

\subsubsection{Comparison with other methods}
\label{subsubsec:recCS_comp}
Since the data are not completely available, another concept is to utilize matrix completion method to interpolate the data and then apply a BSS method on the interpolated data. Therefore, in this subsection, we will compare DecGMCA with the matrix completion followed by a BSS method. Herein, we use a classical SVT algorithm~\cite{Cai2010} to solve the Matrix Completion (MC) problem and GMCA to solve the BSS problem on the interpolated data. In the rest of this subsection, we repeat the same simulation of DecGMCA but utilize MC+GMCA and we compare their performances. For all the figures in this subsection, solid curves represent medians of criteria by applying DecGMCA while dashed curves represent medians of criteria by applying MC+GMCA.

\begin{enumerate}
\item[-] \textit{Sub-sampling effect:} Similarly, we study firstly the sub-sampling effect. In \cref{fig:resCS_MC:a}, we can see the performance of MC+GMCA decreases dramatically when the mask degrades. This might be likely when the loss of information becomes severe, the low-rank hypothesis is no longer valid. As a result, the matrix completion cannot correctly interpolate the data, which means that the following source separation procedure performs badly. Comparing the performance of DecGMCA with MC+GMCA at its turning points (30\% and 60\% for the number of channels 20 and 10 respectively), we can see that DecGMCA conserves well the continuity of both criteria and outperforms MC+GMCA even when the mask is very bad-conditioned. One should notice that when mask is relatively good, DecGMCA still outperforms MC+GMCA. This is due to the fact that DecGMCA takes all of the data into account and simultaneously processes source separation and sub-sampling effect, while MC+GMCA considers them separately. Consequently, the blind source separation in MC+GMCA relies on the quality of matrix completion, which in fact approximates the data interpolation and produces a negligible bias. { Interestingly, the separation performances of the DecGMCA seem to degrade when the average number of available measurements per frequency in the Fourier domain ({\it i.e.} the product of the sub-sampling ratio and the total number of observations) is roughly of the order of the number of sources. In that case, the resulting problem is close to an under-determined BSS problem. In that case the identifiability of the sources is not guaranteed unless additional assumptions about the sources are made. In this setting, it is customary to assume that the sources have disjoint supports in the sparse domain, which is not a valid assumption in the present paper. Additionally, radio-interferometric measurements are generally composed of a large amount of observations for few sources to be retrieved. Furthermore, in contrast to the fully random masks we considered in these experiments, real interferometric masks exhibit a denser amount of data at low frequency and their evolution across channels is mainly a dilation of the sampling mask in the Fourier domain. This entails that the sampling process across wavelegengths is highly correlated, which is a more favorable setting for blind source separation. Altogether, these different points highly mitigate the limitations of the DecGMCA algorithm due to sub-sampling in a realistic inferometric imaging setting.}

\begin{figure}[ht!]
	\begin{center}
	\begin{tabular}{c}
		\subfigure[Performance of DecGMCA in terms of resolution ratio. The number of sources is 5 and the SNR is 60 dB. Abscissa: resolution ratio between best resolved PSF and worst resolved PSF. Ordinate: criterion of mixing matrix for left figure and SDR for right figure.]{\includegraphics[width=0.8\textwidth]{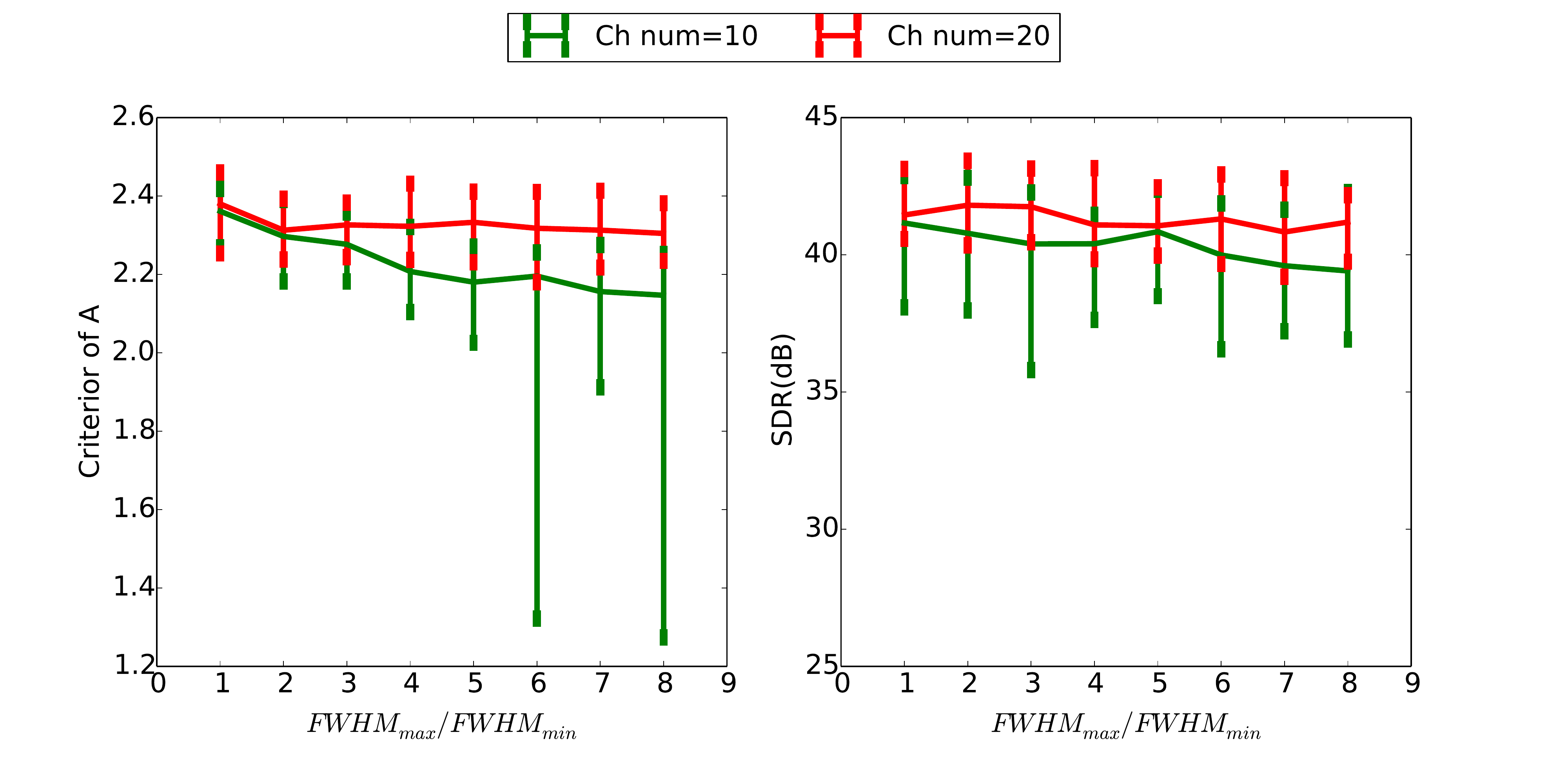}\label{fig:resDec:a}} \\ 
		\subfigure[Performance of DecGMCA in terms of the number of sources. The resolution ratio is 3 and the SNR is 60 dB. Abscissa: number of sources. Ordinate: criterion of mixing matrix for left figure and SDR for right figure.]{\includegraphics[width=0.8\textwidth]{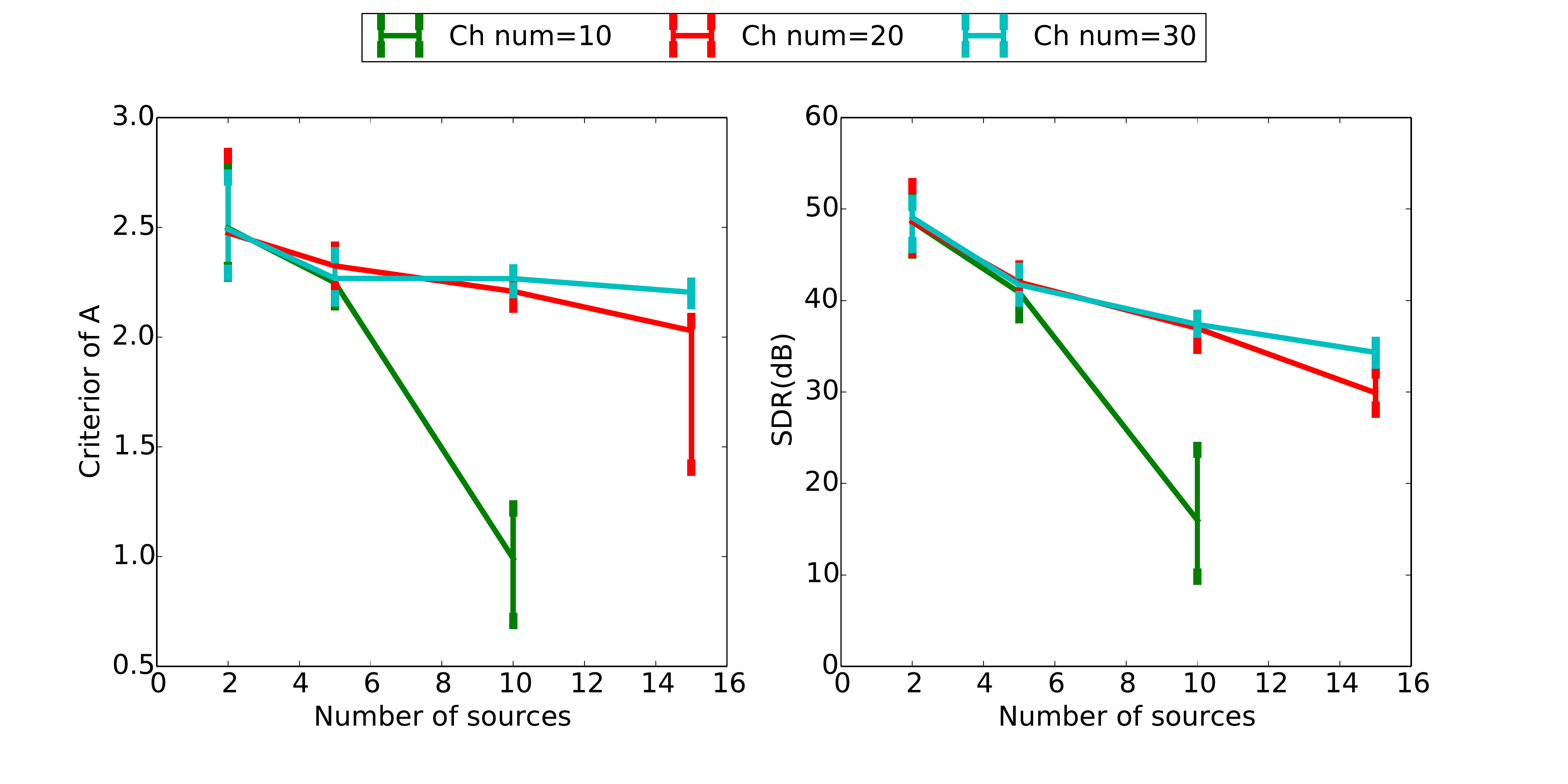}\label{fig:resDec:b}} \\
		\subfigure[Performance of DecGMCA in terms of SNR. The resolution ratio is 3 and the number of sources is 5. Abscissa: SNR. Ordinate: criterion of mixing matrix for left figure and SDR for right figure.]{\includegraphics[width=0.8\textwidth]{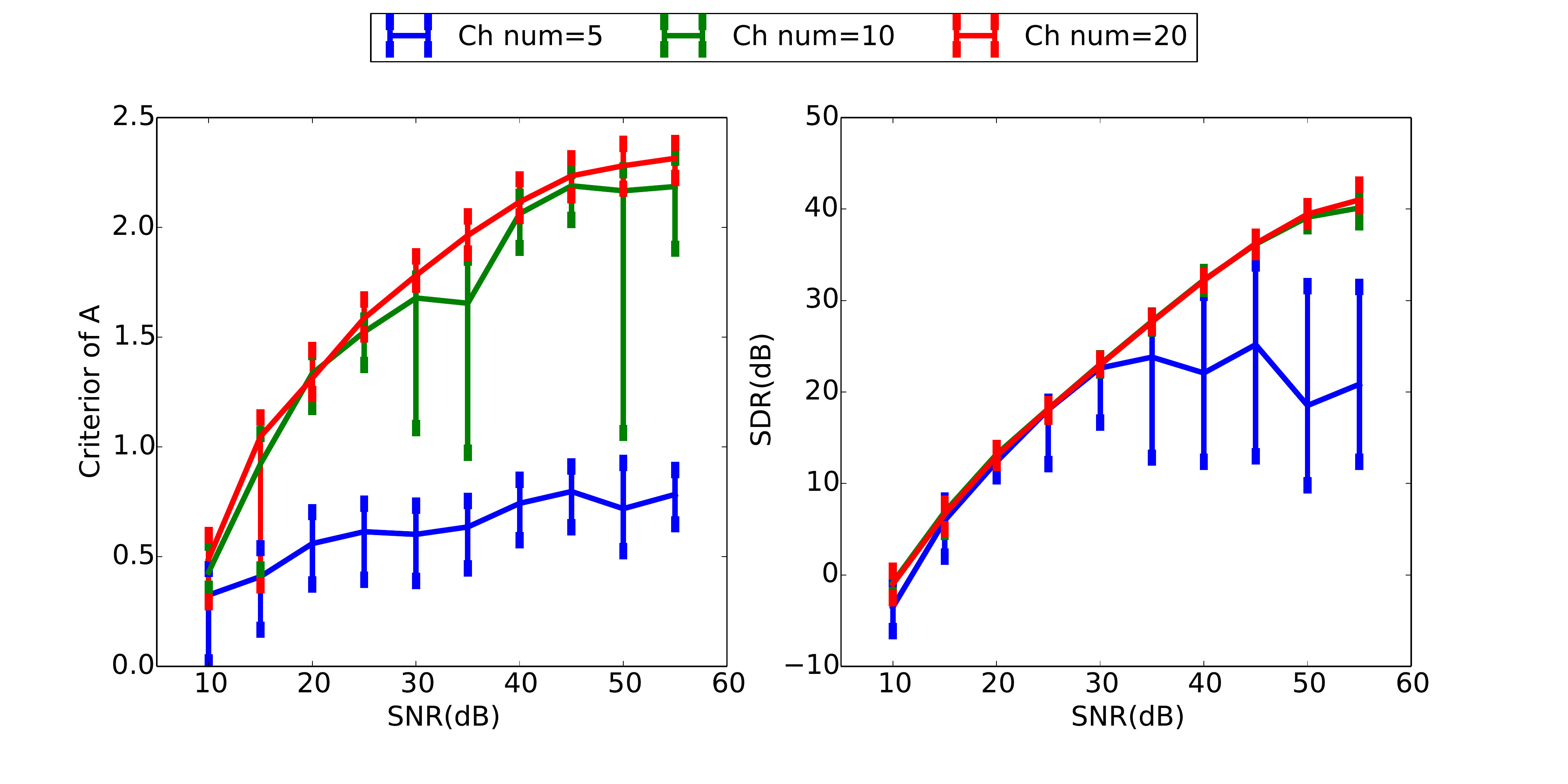}\label{fig:resDec:c}}
	\end{tabular}
	\end{center}
\caption{Multichannel deconvolution BSS simulation (1): study of DecGMCA. The parameters are the resolution ratio, the number of sources and the SNR from top to bottom. The curves are the medians of 50 realizations and the errorbars corresponds to 60\% of ordered data around the median. Blue curves, green curves, red curves and cyan curves represent the number of channels=5, 10, 20 and 30 respectively.}
\label{fig:resDec}
\end{figure}
\clearpage

\item[-] \textit{Number of sources:} Considering the number of sources, in \cref{fig:resCS_MC:b}, we can observe that the performance of MC+GMCA decreases more rapidly than DecGMCA. It can be explained by the fact that a large number of sources complicates the matrix completion (MC) step. Thus, the interpolated data is biased, which affects the following source separation (GMCA). Conversely, as for DecGMCA, jointly solving the data interpolation and the source separation avoids the bias and gives better results.
 
\item[-] \textit{SNR:} In terms of SNR, in \cref{fig:resCS_MC:c}, both DecGMCA and MC+GMCA are stable. However, DecGMCA still outperforms MC+GMCA. One should remark that in MC+GMCA the noise affects both the MC and GMCA steps, which makes the results less accurate. In DecGMCA, as we integrate the demasking in the source separation, we reduce the bias produced by the noise. Hence, DecGMCA has more tolerance to noisy data. {  The DecGMCA tends to perform correctly for large values of the SNR ({\it i.e.} typically larger than $40$ dB) but rapidly fails at low SNR. As we mentionned before, the experimental setting we considered in the present paper is more conservative than realistic inferometric settings. Indeed, radio-interferometric data are generally made of hundreds of measurements, which will dramatically help improving the performances of the DecGMCA algorithm in the low SNR regime.}.
\end{enumerate}

\subsection{Multichannel deconvolution and BSS simulation}
\label{subsec:simDec}
Then, as for the general form of the operator $\hat{\mathbf{H}}$, we will extend the above experiments to a more general multichannel deconvolution case. In practice, because of instrumental limits, the resolution of the PSF of multichannel sensors is dependent on the channel. We assume that the resolution of the Gaussian-like PSF increases linearly along the channel index. The PSF function in channel $\nu$ is defined by: $\text{F}_\nu(x)=\exp(\frac{x^2}{2\sigma_{\nu}^2})$, where the coordinate in Fourier space $x\in [-2047,2048]$. As FWHM (Full Width at Half Maximum, which is conceptually considered as a standard deviation) is commonly used to define the resolution of the PSF, we define a resolution ratio based on FWHM to quantify the variation of the PSF hereafter: $\text{ratio}=\frac{\text{FWHM}_{max}}{\text{FWHM}_{min}}$. In this section, the best resolved Gaussian-like PSF is always fixed with $\sigma_{max}=1800$. We will utilize our DecGMCA method to realize the multichannel deconvolution BSS. We will study the performance in three parts as well: resolution ratio, the number of sources and SNR. The initial regularization parameter $\epsilon^{(0)}$ is set to be 1 and $\epsilon$ decreases to $10^{-5}$ for all the experiments in this section.

\subsubsection{On the study of DecGMCA}
\label{subsec:recDec}
\begin{enumerate}
\item[-]\textit{Resolution ratio:} The first parameter is the resolution ratio. In this paragraph, we define different resolution ratios to study the performance of DecGMCA. The resolution ratio is varied from 1 to 8. Therefore, ratio=1 means all channels have the same resolution, while ratio=8 means the largest difference between the best resolution and the worst resolution. $\text{N}_\text{s}$ is fixed to 5 and the SNR is fixed to 60 dB. $\text{N}_\text{c}$ is set to 10 and 20. In \cref{fig:resDec_dec:a}, we can observe that if $\text{N}_\text{s}=10$, the performance of DecGMCA becomes unstable as resolution ratio increases. This means that when the resolution ratio becomes large, the system becomes ill-conditioned and the noise is likely to explode. However, if we increase the number of channels to 20, both criteria become more stable. Indeed, as the number of channels increases, we have more information to help to estimate $\mathbf{A}$ and $\mathbf{S}$, yielding more accurate estimates. One might notice that the SDR does not change no matter which resolution ratio or number of channels is used. This can be interpreted by the fact that even though the system is ill-conditioned, DecGMCA can successfully regularize the ill-conditioned PSF. Although the $\Delta_\mathbf{A}$ becomes unstable when the resolution ratio becomes large, its median is of good precision.

\item[-] \textit{Number of sources:} The second parameter is the number of sources. In this paragraph, $\text{N}_\text{s}$ is set to 2, 5, 10 and 15. The resolution ratio is 3 and the SNR is 60 dB. $\text{N}_\text{c}$ is set to 10, 20 and 30. In \cref{fig:resDec:b}, we can observe that when $\text{N}_\text{c}$ is 2 and 5, all criteria are very good ($\Delta_\mathbf{A}>2$ and $\text{SDR}>40$ dB) and almost superposed. This is due to the fact that the number of channels is always sufficiently large compared to the number of sources and the ill-conditioned PSF is not difficult to be inverted. As expected, when the number of sources increases, the system becomes more complicated and both criteria decrease. In particular, considering the most difficult case in our test ($\text{N}_\text{c}=10$ for $\text{N}_\text{s}=10$), DecGMCA does a poor job at regularizing the system and the effect of ill-conditioned PSF significantly degrades both criteria.

\item[-] \textit{SNR:} In the end, concerning the impact of noise on DecGMCA, the SNR is varied from 10 dB to 55 dB. The resolution ratio is fixed to 3 and $\text{N}_\text{s}$ is fixed to 5. $\text{N}_\text{c}$ is set to 5, 10 and 20. \Cref{fig:resDec:c} features the evolution of both criteria in terms of SNR. Firstly, when the number of channels is 5, the performance of DecGMCA is always poor. The reason is that recovering 5 sources from 5 channels is the limit of the BSS problem, but the ill-conditioned PSF raises the difficulty. Even though DecGMCA can regularize the system, its effectiveness is limited when SNR is very low. When $\text{N}_\text{c}$ is 10 or 20, since more observations are available, both criteria grow rapidly along with the SNR. 
When the SNR is low, the data is so noisy that even with regularization DecGMCA cannot efficiently select useful information. Conversely, when SNR is high, especially above 40 dB, DecGMCA is able to accurately estimate the mixing matrix and the sources. One might notice that generally $\Delta_\mathbf{A}$ is more unstable than SDR. It means that the criterion of $\mathbf{A}$ is more sensitive to the noise.
\end{enumerate} 

\subsubsection{Comparison with other methods}
\label{subsubsec:recDec_comp}
DecGMCA considers BSS and deconvolution simutaneously and naturally it gives a better result than considering them separately. In order to validate, we compare DecGMCA with different approaches:
\begin{enumerate}
\item[$\bullet$] Blind Source Separation only, without deconvolution. GMCA is used for BSS.
\item[$\bullet$] a channel by channel deconvolution using ForWaRD followed by BSS (ForWaRD+GMCA)
\end{enumerate}

In the rest of this subsection, we utilize GMCA and ForWaRD+GMCA to repeat our DecGMCA simulation and compare their performances. For all the figures in this subsection, solid curves represent medians of criteria by applying DecGMCA while pointed curves and dashed curves represent medians of criteria by only GMCA and ForWaRD+GMCA respectively.

\begin{figure}[ht!]
	\begin{center}
	\begin{tabular}{c}
		\subfigure[Comparison among DecGMCA, GMCA and ForWaRD+GMCA in terms of resolution ratio. The number of sources is 5 and the SNR is 60 dB. Abscissa: ratio between the best resolved PSF and the worst resolved PSF. Ordinate: criterion of mixing matrix for left figure and SDR for right figure.]{\includegraphics[width=0.7\textwidth]{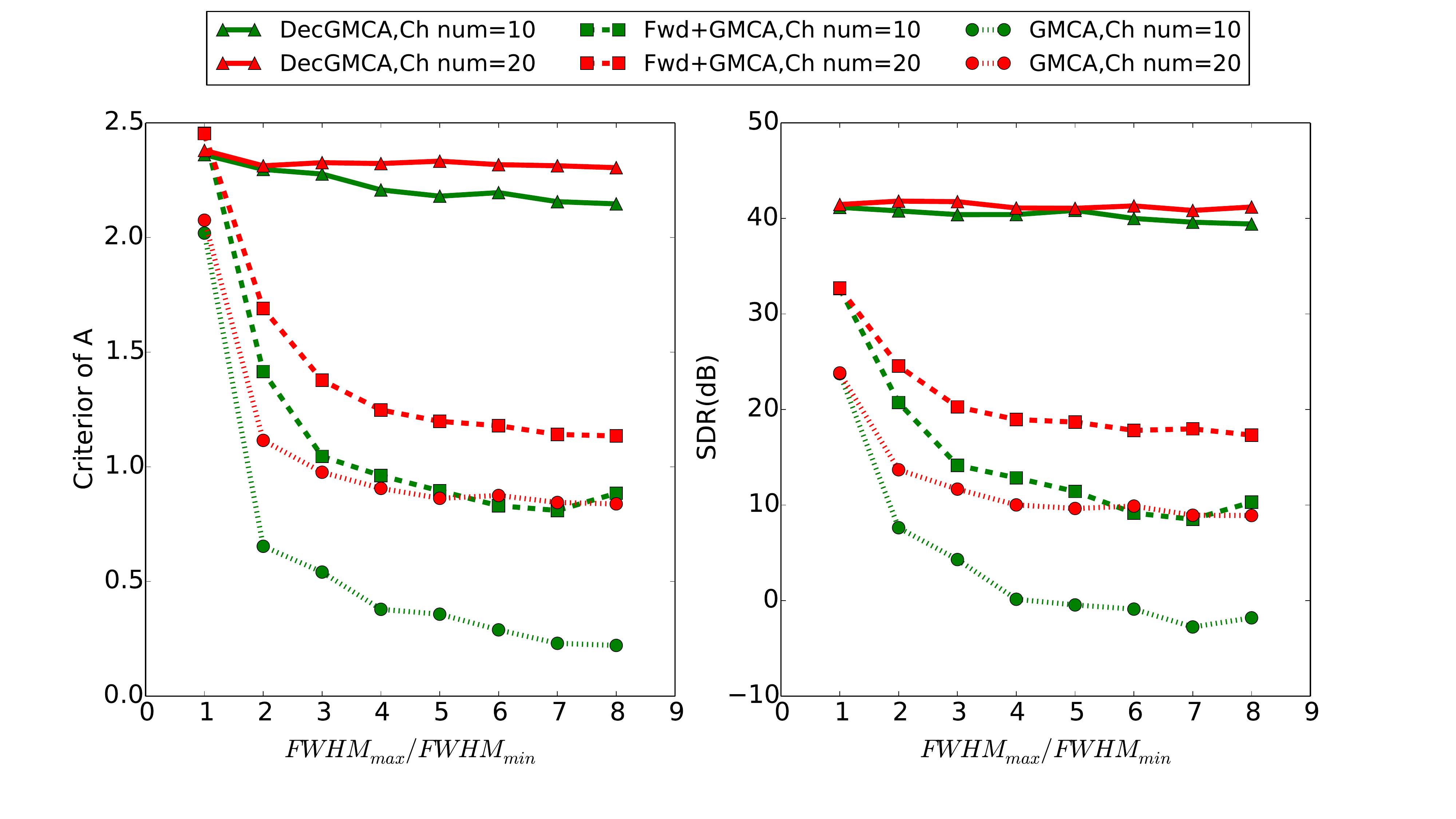}\label{fig:resDec_dec:a}} \\ 
		\subfigure[Comparison among DecGMCA, GMCA and ForWaRD+GMCA in terms of the number of sources. The resolution ratio is 3 and the SNR is 60 dB. Abscissa: number of sources. Ordinate: criterion of mixing matrix for left figure and SDR for right figure.]{\includegraphics[width=0.7\textwidth]{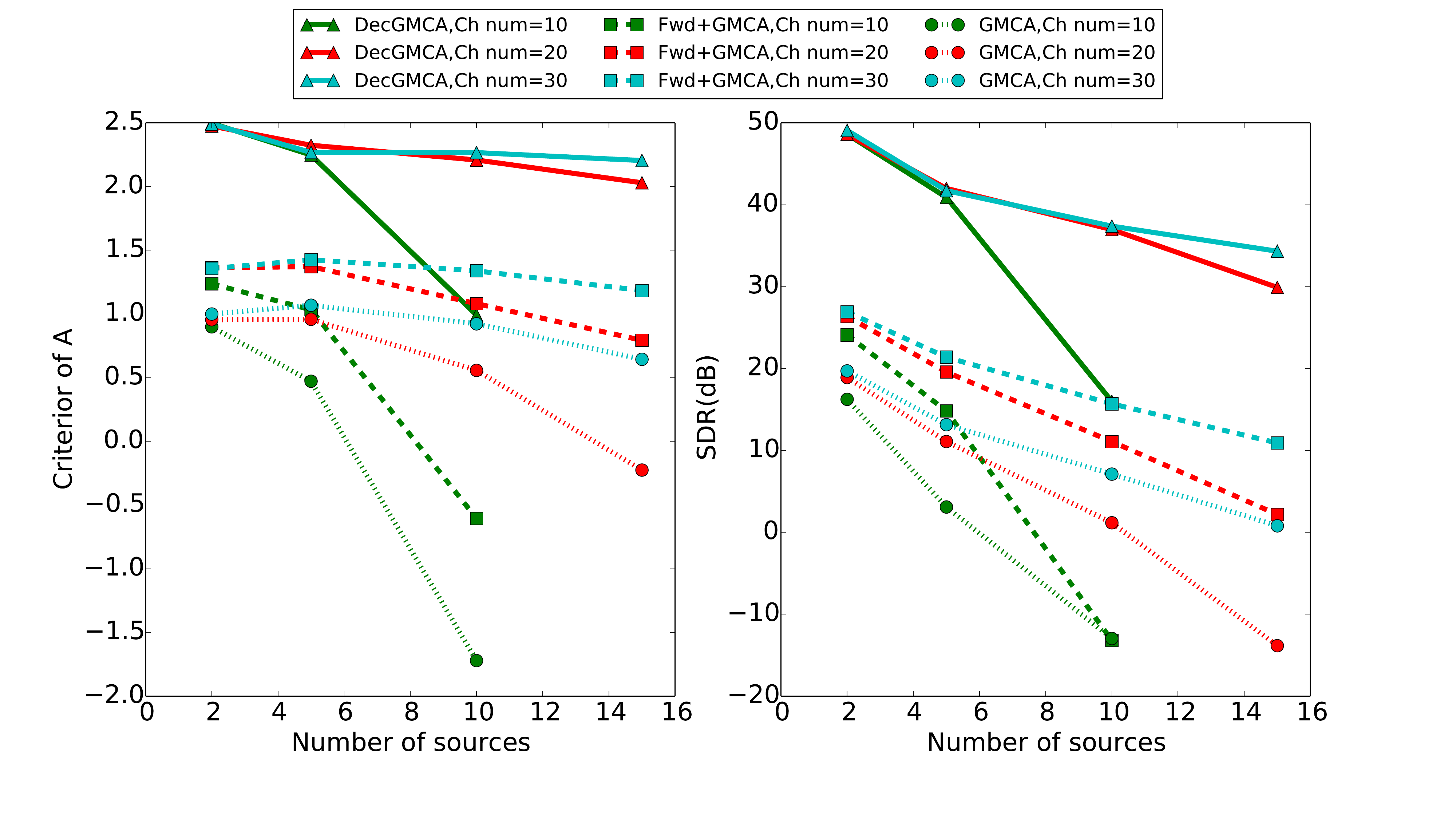}\label{fig:resDec_dec:b}} \\
		\subfigure[Comparison among DecGMCA, GMCA and ForWaRD+GMCA in terms of SNR. The resolution ratio is 3 and the number of sources is 5. Abscissa: SNR. Ordinate: criterion of mixing matrix for left figure and SDR for right figure.]{\includegraphics[width=0.7\textwidth]{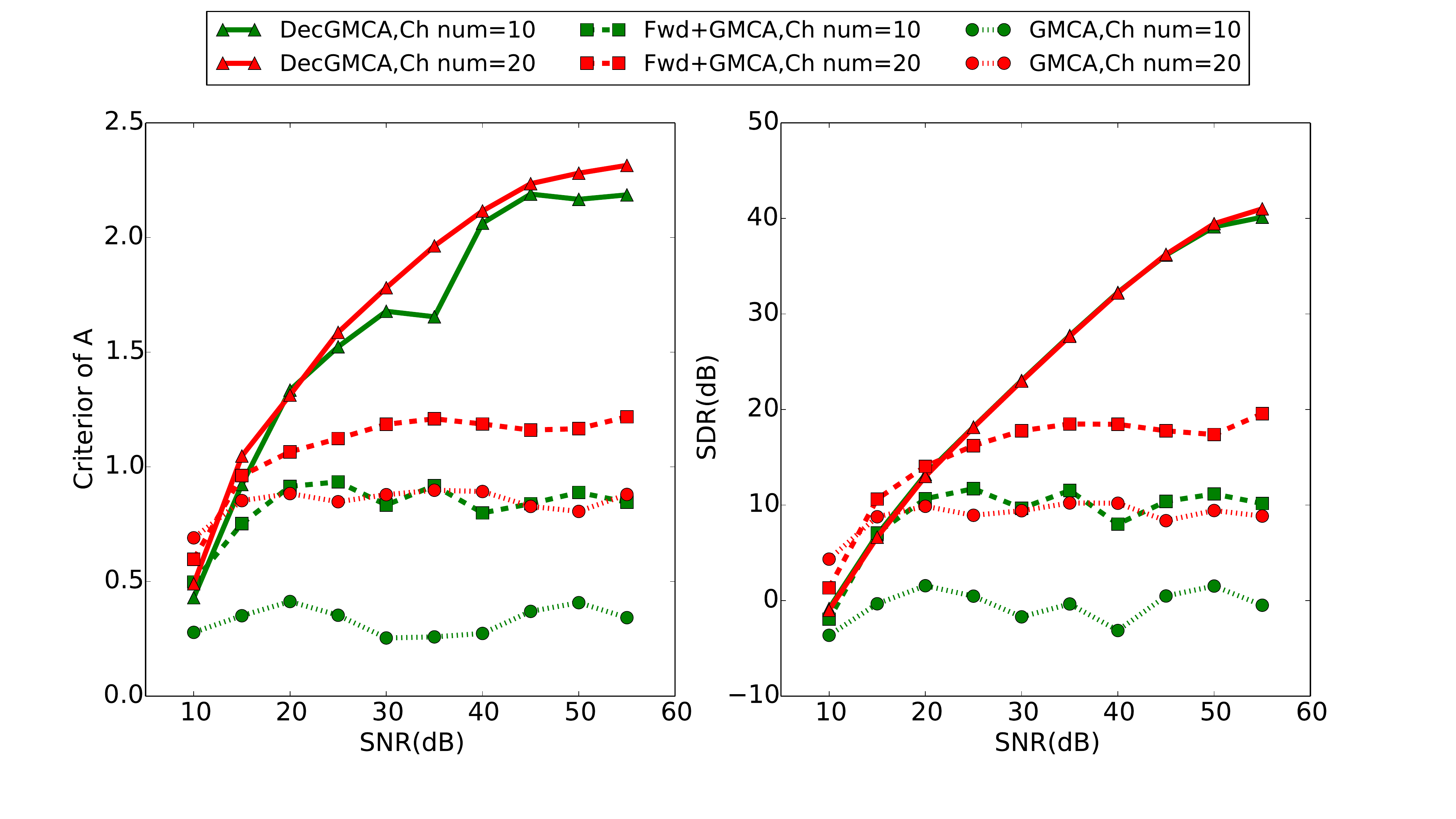}\label{fig:resDec_dec:c}}
	\end{tabular}
	\end{center}
\caption{Multichannel deconvlution and BSS simulation (2): comparison among DecGMCA (joint deconvolution and BSS), only GMCA (BSS) and ForWaRD+GMCA (channel by channel deconvolution followed by BSS). The parameters are the resolution ratio, the number of sources and the SNR from top to bottom. The curves are the medians of 50 realizations. Blue curves, green curves, red curves and cyan curves represent the number of channels=5, 10, 20 and 30 respectively.}
\label{fig:resDec_dec}
\end{figure}

\begin{enumerate}
\item[-]\textit{Resolution ratio:} In terms of resolution ratio, \cref{fig:resDec_dec:a} displays the performance of DecGMCA, GMCA and ForWaRD+GMCA. As GMCA does not consider the varied PSFs and does not perform the deconvolution, GMCA provides the worst results. Although ForWaRD+GMCA takes into account the deconvolution, it processes the deconvolution on channel by channel instead of the whole dataset. Thus, it neglects the correlation between channels and gives a biased deconvolution, leading to a much worse result than DecGMCA. It is also interesting to notice that when the resolution ratio is small, the difference between the three methods is smaller than when the resolution is large. This is because the PSFs are more varied when the resolution ratio is larger. DecGMCA, which simultaneously considers deconvolution and BSS, is able to capture the change of different PSFs and adapt to the complex data. Therefore, DecGMCA outperforms the others when the PSFs are widely varied.

\item[-] \textit{Number of sources:} Concerning the number of sources, from \cref{fig:resDec_dec:b}, we can see that as the number of sources increases, the system becomes more complicated and the criteria of all of three methods degrade. Yet, DecGMCA always outperforms ForWaRD+GMCA by at least 25 dB in SDR. This is because when considering simultaneous BSS and deconvolution, much more global information can be conserved and the solution is less biased. Besides, among three methods, GMCA is again the worst one as GMCA neglects the blurring effect caused by the PSF.

\item[-] \textit{SNR:} Finally, in terms of SNR, in \cref{fig:resDec_dec:c}, we can see the performance of GMCA is very poor. This is not only the impact of the noise but also the neglect of the blurring effect. As for ForWaRD+GMCA, both criteria grow slightly in terms of SNR (when $\text{SNR}<20$ dB) as it takes the blurring effect into account and restores sources, but ForWaRD+GMCA attains the saturation point rapidly as it cannot perform better with channel by channel deconvolution. However, the performance of DecGMCA grows rapidly as SNR increases because simultaneously considering deconvolution and BSS benefits from the data for the recovery of the mixing matrix and sources.
\end{enumerate} 

\clearpage

\section{Application to astrophysical data}
\label{sec:experiments2}
In astrophysics, sources are often Gaussian-like. The left column of \cref{fig:illstr_S} displays three astrophysical sources. It has been shown that the starlet dictionary \cite{starck2002} is the best representation for such isotropic sources. In spectroscopy, the astrophysical source has a characteristic spectrum $f(x)\propto x^{-k}$, which generally respects power law with a specific index. Through interferometers, we can capture these sources and study their spectra. However, the problem of interferometry imaging is that the observation is sub-sampled in Fourier space. Besides the sub-sampling effect, the PSF, or the angular resolution of the interferometer, is limited by its beamforming and varies as a function of wavelength. Therefore, we extend the numerical experiments in the previous section to the case where the operator $\hat{\mathbf{H}}$ takes not only the sub-sampling effect but also the blurring effect into account. 

\begin{figure}[ht!]
	\begin{center}
	\begin{tabular}{c c}
		\subfigure[Example of the best resolved PSF over total 20 channels.]{\includegraphics[width=0.35\textwidth]{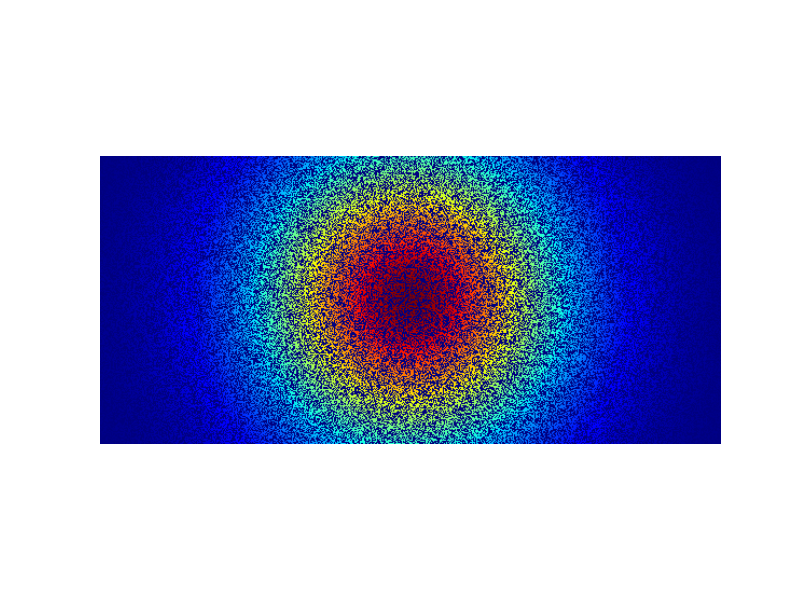}} & \subfigure[Example of the worst resolved PSF over total 20 channels.]{\includegraphics[width=0.35\textwidth]{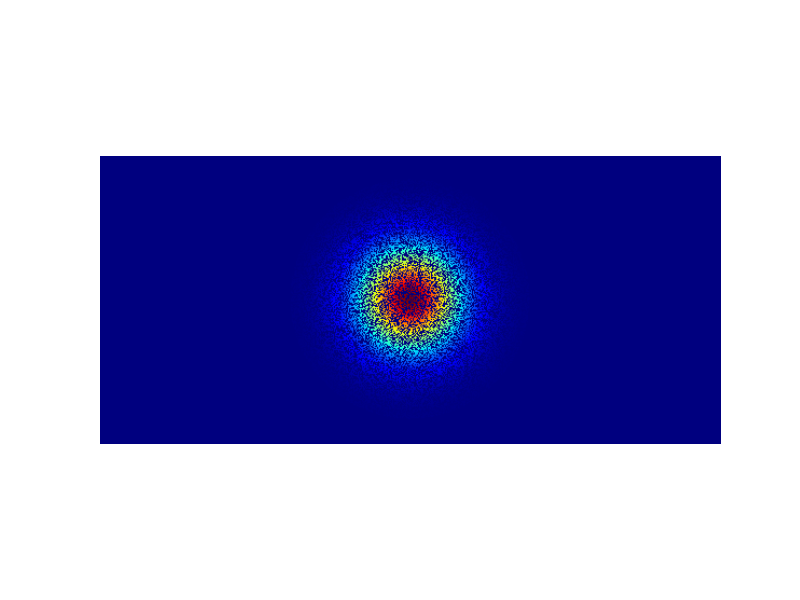}} \\
	\end{tabular}
	\end{center}
\caption{Illustration of masked PSFs (in Fourier space): the resolution ratio is 3 and the percentage of active data is 50\%).}
\label{fig:illstr_kern}
\end{figure}

\begin{figure}[h!]
\centering
\includegraphics[width=0.35\textwidth]{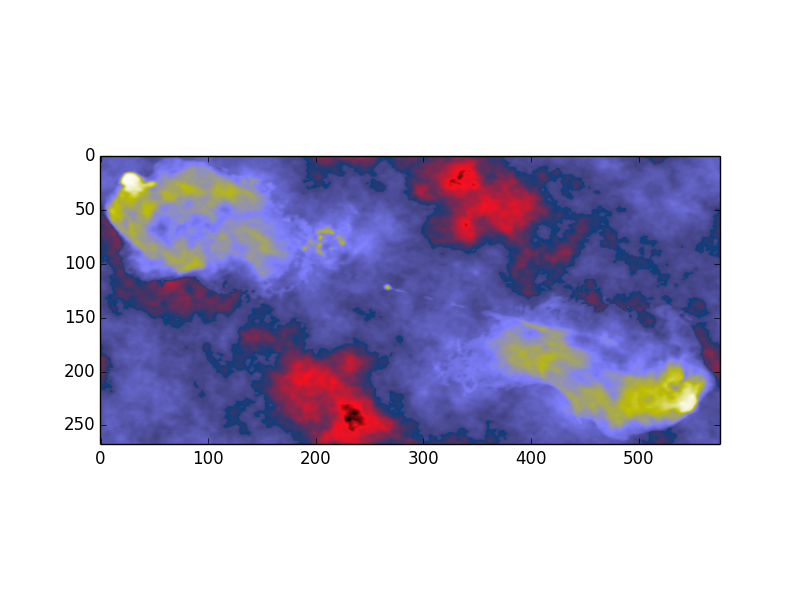}
\caption{Illustration of 1 out of 20 mixtures blurred by the masked PSFs and contaminated by the noise: the resolution ration is 3, the percentage of active data is 50\% and the SNR is 60 dB. Remark: The real data are in Fourier space, but the data here is transformed to image space for better visualization.}
\label{fig:illstr_data}
\end{figure}

For simplicity, we assume that the number of observation channels is 20. The resolution ratio of the best resolved PSF and the worst resolved PSF is 3 and the percentage of active data in Fourier space is 50\%. In addition, the noise level is fixed to 60 dB. \cref{fig:illstr_kern} illustrates two masked PSFs (the best resolved one and the worst resolved one) in Fourier space and \cref{fig:illstr_data} gives an example of 1 mixture out of 20. We can see that sources are drown in the ``dirty'' image and mixed with each other. It seems to be very challenging to discriminate and recover these sources from such an ill-conditioned PSF.

\begin{figure}[ht!]
\centering
\includegraphics[width=0.95\textwidth]{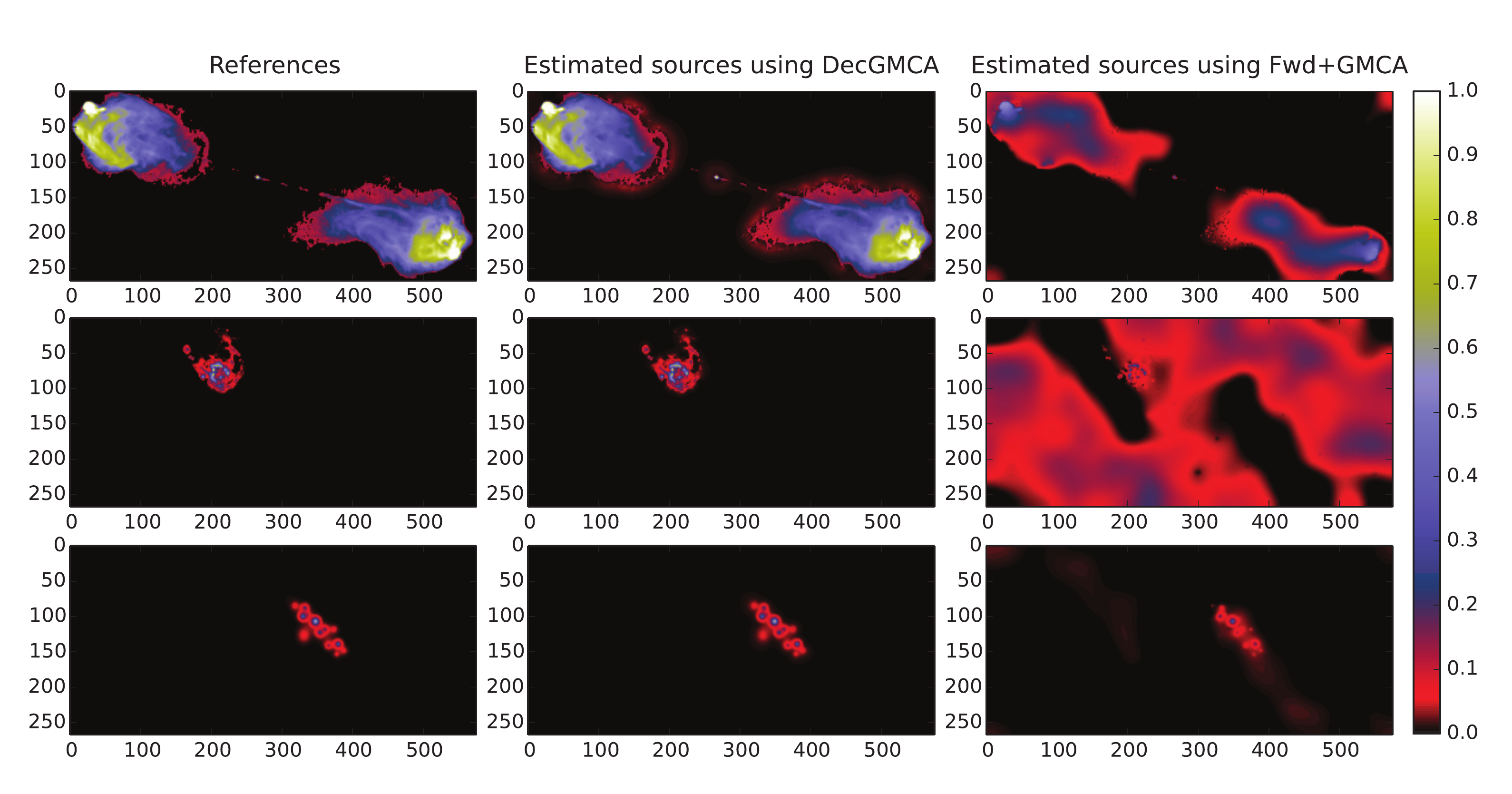}
\caption{Illustration of DecGMCA applied on astrophysical images. The raw data is blurred by the masked PSFs and contaminated by the noise: the resolution of PSF is linearly declined along 20 channels with resolution ratio=3, besides, PSFs are masked with percentage of active data=50\% and the SNR is 60 dB. \cref{fig:illstr_data} shows an example of the raw data in image domain. We apply DecGMCA to separate and recover sources. Left column: Ground-truth of three sources from top to bottom. Middle column: Estimate of three sources by using DecGMCA from top to bottom. Right column: Estimate of three sources by using ForWaRD+GMCA from top to bottom.}
\label{fig:illstr_S}
\end{figure}

We set the wavelet scale to 5, final threshold level to $3\sigma$ ($\sigma$ stands for the standard deviation of the noise) and the regularization parameter $\epsilon$ is initialized as $10^{-5}$ and decreases to $10^{-6}$. Having applied DecGMCA on such ``dirty'' data, we can see recovered sources presented in the middle column of \cref{fig:illstr_S}. The sources are well deconvolved and separated. Visually, compared to the references in the left column of \cref{fig:illstr_S}, the main structures of sources are well positioned and even the detailed features of estimated sources are successfully recovered. This means that the estimated sources by using DecGMCA have a very good agreement with the references. However, if we first applied the ForWaRD method to perform the channel by channel deconvolution and then applied GMCA method to separate sources, the results would not be reliable. We can see in the right column of \cref{fig:illstr_S} that the sources cannot be recovered properly. The main structure of the first source is recovered but not well deconvolved and the structure is biased; the second source cannot even be recovered with many artifacts present in the background; the third source is successfully separated but the structures are biased and not compact.

\begin{figure}[h!]
	\begin{center}
	\begin{tabular}{c c}
		\subfigure[Residuals between estimated sources by using DecGMCA and ground-truth sources, the relative errors are 0.14\%, 0.27\% and 0.36\% from top to bottom.]{\includegraphics[width=0.45\textwidth]{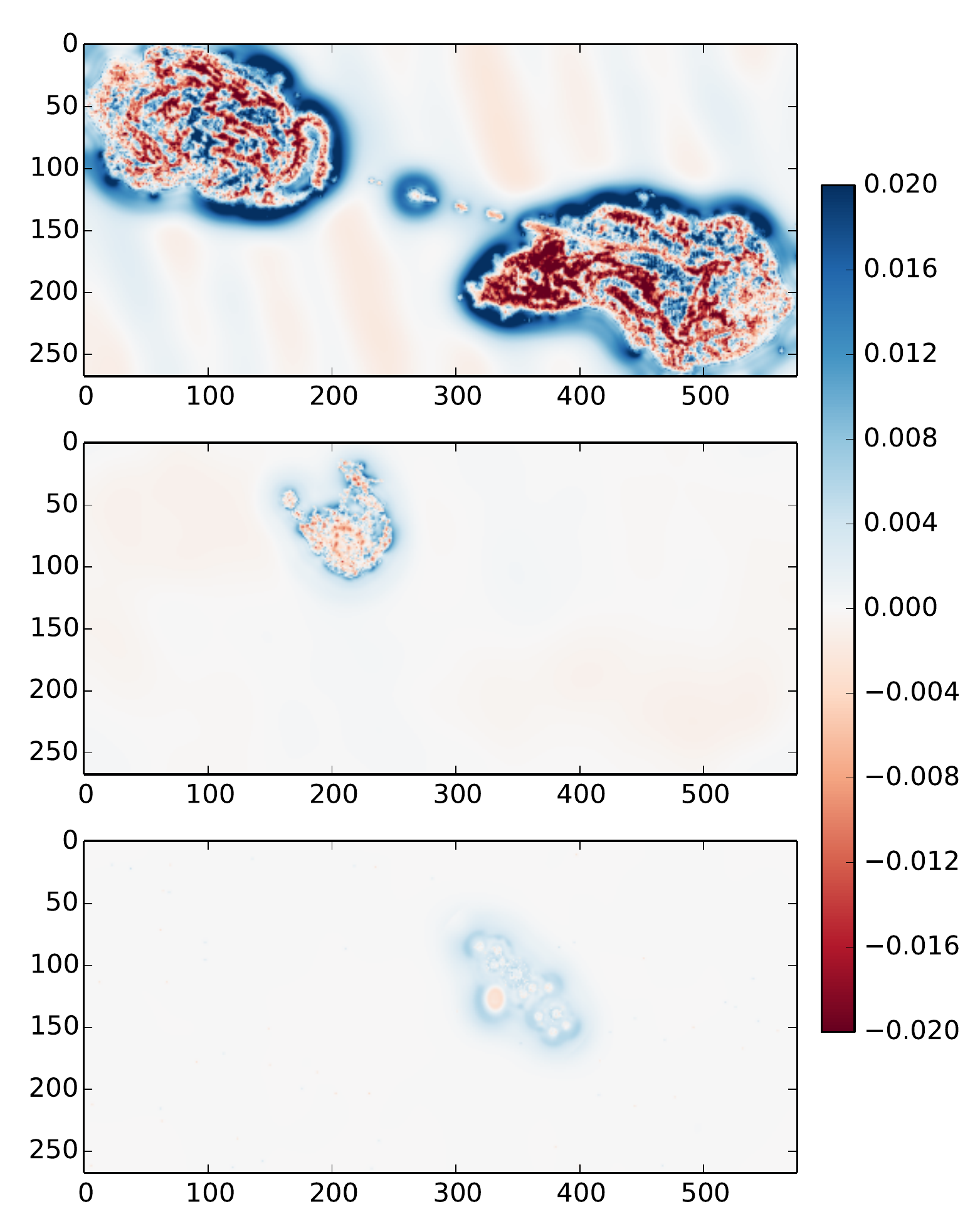}\label{fig:illstr_res:a}} &
		\subfigure[Residuals between estimated sources by using ForWaRD+GMCA and ground-truth sources, the relative errors are 54.74\%, 1279.21\% and 30.12\% from top to bottom.]{\includegraphics[width=0.45\textwidth]{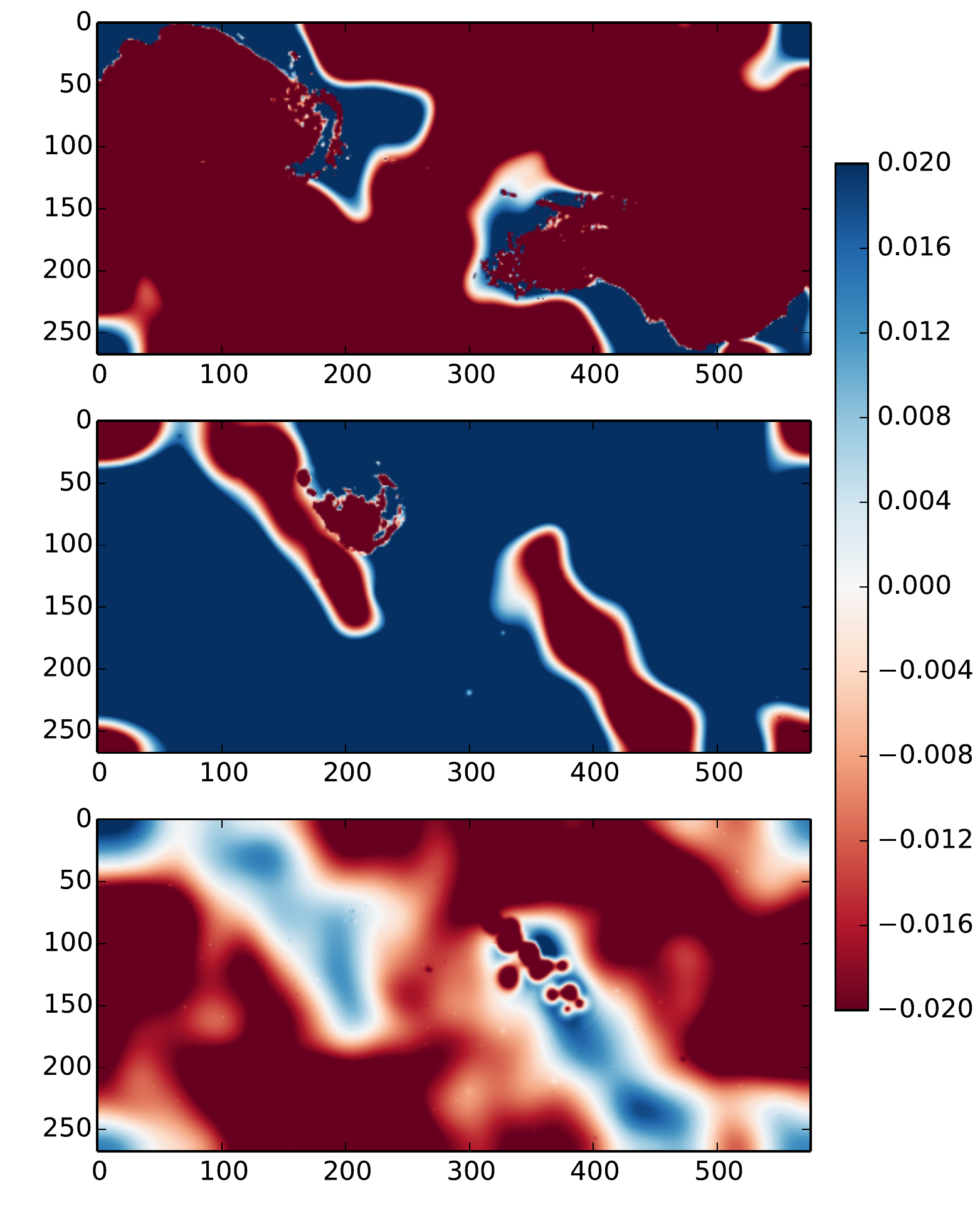}\label{fig:illstr_res:b}}
	\end{tabular}
	\end{center}
\caption{Comparison between joint deconvolution and BSS using DecGMCA and channel by channel deconvolution followed by BSS using ForWaRD+GMCA. Left column: residuals between estimated sources and ground-truth using DecGMCA. Right column: residuals between estimated sources and ground-truth using ForWaRD+GMCA.}
\label{fig:illstr_res}
\end{figure}

Furthermore, by computing residuals between estimated sources and ground-truth sources, \cref{fig:illstr_res} displays the error map of DecGMCA (left column) and ForWaRD+GMCA (right column). We also compare their relative errors which are shown in \cref{tab:res_re}. We can see that DecGMCA is very accurate with the relative errors for three sources 0.14\%, 0.27\% and 0.36\% respectively, which shows that our estimated sources have a good agreement with the ground-truth. However, as for the ForWaRD+GMCA, since sources are not cleanly separated and recovered, the residuals are significant. The relative errors are tremendous, respectively 54.74\%, 1279.21\% and 30.12\% respectively.

\begin{table}[ht!]
\caption{Comparison of relative errors between DecGMCA and ForWaRD+GMCA}
\label{tab:res_re}
\centering
\begin{tabular}{|c|c|c|}\hline
Sources & DecGMCA & ForWaRD+GMCA \\ \hline
1 & 0.14\% & 54.74\% \\ 
2 & 0.27\% & 1279.21\% \\
3 & 0.36\% & 30.12\% \\ \hline
\end{tabular}
\end{table}

\section{Software}
\label{sec:software}
In order to reproduce all the experiments, the codes used to generate the plots presented in this paper will be available online at \url{http://www.cosmostat.org/software/gmcalab/}

\section{Conclusions}
\label{sec:conclusions}
In this article, we investigated an innovative solution to the DBSS problems, where deconvolution and blind source separation need to be solved simultaneously. The proposed algorithm, dubbed Deconvolution GMCA (DecGMCA) builds upon sparse signal modeling and a novel projected least-squares algorithm to solve BSS problems from incomplete measurements and/or blurred data. Numerical experiments, in the multichannel compressed sensing and multichannel deconvolution settings, have been carried out that show the efficiency of the proposed DecGMCA algorithm. These results further emphasize the advantage of the joint resolution of both the deconvolution and unmixing problems rather than an independent processing. DecGMCA has been applied to astrophysical data, which illustrates that it is a very good candidate for processing radioastronomy data. Future work will focus on extending the proposed approach to account for more complex acquisition models.
\newpage

\section*{Acknowledgments}
This work is supported by the CEA DRF impulsion project COSMIC and the European Community through the grants PHySIS (contract no. 60174), DEDALE (contract no. 665044) and LENA (contract no. 678282) within the H2020 Framework Programe. The authors would like to thank Samuel Farrens for useful comments.

\appendix
\section{Proximal algorithms to solve \texorpdfstring{\Cref{eq:decgmca_S}}{Lg}} Step \ref{algo:line28} in \Cref{alg:DecGMCA} consists in using the Condat-V\~{u}~\cite{vu2013,condat2013} algorithm to improve the estimate of $\mathbf{S}$ with respect to $\mathbf{A}$. As we formulate the sub-problem \cref{eq:decgmca_S} in an analysis framework, the proximal operator $||\cdot\mathbf{\Phi}^t||_p$ ($p=0$ or $1$) is not explicit. The advantage of Condat-V\~{u} or other primal-dual algorithms is that we do not need an inner loop to approach the proximal operator as done in the Forward-Backward algorithm~\cite{combettes2005}. The detailed algorithm is presented in \Cref{alg:condat-vu}.

\begin{algorithm}[ht!]
\caption{Condat-V\~{u} algorithm}
\label{alg:condat-vu}
\begin{algorithmic}[1]
\STATE{\textbf{Input}: Observation $\hat{\mathbf{Y}}$, operator $\hat{\mathbf{H}}$, mixing matrix $\mathbf{A}$, maximum iterations $\text{N}_\text{i}$, threshold $\lambda$,$\tau>0$,$\eta>0$}
\STATE{Initialize $\mathbf{S}^{(0)}$ as the last estimate by using ALS scheme,$\boldsymbol\alpha^{(0)}=\mathbf{S}\mathbf{\Phi}$}
\FOR{$i=1,\cdots,\text{N}_\text{i}$}
\STATE{$\bullet$ Obtain sources in Fourier space by FFT}
\STATE{$\hat{\mathbf{S}}^{(i)}=\text{FT}(\mathbf{S}^{(i)})$}
\STATE{$\bullet$ Compute the residual in Fourier space}
\STATE{$\mathbf{R}^{(i)}=\hat{\mathbf{Y}}-\hat{\mathbf{H}}\odot\mathbf{A}\hat{\mathbf{S}}^{(i)}$}
\STATE{$\bullet$ Update $\mathbf{S}$}
\STATE{$\mathbf{S}^{(i+1)}=\mathbf{S}^{(i)}-\tau\boldsymbol\alpha^{(i)}\mathbf{\Phi}+\tau\text{Re}\left(\text{FT}^{-1}\left(\mathbf{A}^t\left(\hat{\mathbf{H}}^*\odot\mathbf{R}^{(i)}\right)\right)\right)$}
\STATE{$\bullet$ Introduce an intermediate variable}
\STATE{$\textbf{V}^{(i+1)}=2\mathbf{S}^{(i+1)}-\mathbf{S}^{(i)}$}
\FOR{$j=1,\cdots,\text{N}_\text{s}$}
\STATE{$\bullet$ Update $\boldsymbol\alpha$ under sparsity constraint}
\STATE{$\boldsymbol\alpha^{(i+1)}_j=\left(\mathbf{I}-\Th_{\lambda^{(i)}_j}\right)\left(\boldsymbol\alpha^{(i)}_j+\eta\mathbf{V}^{(i+1)}_j\mathbf{\Phi}^t\right)$}
\ENDFOR
\ENDFOR
\RETURN $\mathbf{S}^{(\text{N}_\text{i})}$
\end{algorithmic}
\end{algorithm}

\clearpage

\bibliographystyle{siamplain}
\bibliography{references}
\end{document}